\documentclass{emulateapj}
\usepackage{amsmath}
\usepackage{graphicx}
\shorttitle{PIC simulations of the MRI}
\shortauthors{}

\begin{document}

\title{Local 2D Particle-in-cell simulations of the collisionless MRI}

\author{Mario A. Riquelme and Eliot Quataert}
\affil{Astronomy Department and Theoretical Astrophysics Center, University of California, Berkeley, CA 94720; marh@astro.berkeley.edu, eliot@astro.berkeley.edu}

\author{Prateek Sharma}
\affil{Department of Physics, Indian Institute of Science, India 560012; prateek@physics.iisc.ernet.in}
\and
\author{Anatoly Spitkovsky}
\affil{Department of Astrophysical Sciences, Princeton University; anatoly@astro.princeton.edu}

\begin{abstract} 

  The magnetorotational instability (MRI) is a crucial mechanism of angular momentum transport in a variety of astrophysical accretion disks. In systems accreting at well below the Eddington rate, such as the central black hole in the Milky Way (Sgr A*), the rate of Coulomb collisions between particles is very small, making the disk evolve essentially as a collisionless plasma. We present a nonlinear study of the collisionless MRI using first-principles particle-in-cell (PIC) plasma simulations.
  In this initial study we focus on local two-dimensional (axisymmetric) simulations, deferring more realistic three-dimensional simulations to future work.  For simulations with net vertical magnetic flux, the MRI continuously amplifies the magnetic field, $\vec{B}$, until the Alfv\'{e}n velocity, $v_A$, is comparable to the speed of light, $c$ (independent of the initial value of $v_A/c$).  This is consistent with the lack of saturation of MRI channel modes in analogous axisymmetric MHD simulations.  The amplification of the magnetic field by the MRI generates a significant pressure anisotropy in the plasma ($p_{\perp} \gtrsim p_{||}$, where $p_{\perp}$ and $p_{||}$ are the plasma pressures perpendicular and parallel to the local magnetic field).  We find that this pressure anisotropy in turn excites mirror modes and that the volume averaged pressure anisotropy remains near the threshold for mirror mode excitation.  Particle energization is due to both reconnection and viscous heating associated with the pressure anisotropy.
  Reconnection produces a distinctive power-law component in the energy distribution function of the particles, indicating the likelihood of non-thermal ion and electron acceleration in collisionless accretion disks.  This has important implications for interpreting the observed emission -- from the radio to the gamma-rays -- of systems such as Sgr A*.

\end{abstract}

\keywords{accretion disks, MRI, kinetic plasma effects}

\section{Introduction}
\label{sec:intro}
Accretion disks are ubiquitous in astrophysics and play a fundamental role in areas as diverse as planet formation, gamma ray bursts (GRBs), and accretion onto supermassive black holes in the centers of galaxies. The accretion of gas in disks requires outward transport of angular momentum, typically assumed to be provided by the magnetorotational instability \cite[MRI;][]{BalbusEtAl91, BalbusEtAl98}. The MRI has been widely studied using MHD simulations. However, in many cases the MHD approach is not directly applicable. When the time scale for electron and ion Coulomb collisions is longer than the inflow time in the disk, the plasma is macroscopically collisionless and MHD breaks down. This is the case in radiatively inefficient accretion flow (RIAF) models, applicable when the accretion rate is less than a few percent of Eddington \citep{Narayan1998}. The low rate of Coulomb collisions implies that ions and electrons are thermally decoupled, so the plasma should be two-temperature. RIAFs are ubiquitous, occurring, for example, in the low-hard state of X-ray binaries \citep[e.g.,][]{EsinEtAl97}, and around the central supermassive black hole in the Milky Way (Sgr A*) and most nearby galaxies. \newline

The first efforts to understand the MRI in the collisionless limit were done using the kinetic MHD approach (Quataert et al. 2002; Sharma et al. 2003; Sharma et al. 2006; see also the closely related work by Balbus 2004 and Islam \& Balbus 2005). These studies highlighted the importance of pressure anisotropies with respect to the local magnetic field, $\vec{B}$, and incorporated the evolution of the pressure parallel and perpendicular to $\vec{B}$ in a fluid model of the plasma. In particular, an increase in the magnetic field due to the MRI causes the plasma temperature $T_j$ (and pressure, $p_j$) of a given particle species $j$ to increase in the direction perpendicular to the local magnetic field due to the conservation of the magnetic moment, $\mu_j$, on scales larger than their Larmor radius ($\mu_j \equiv p_{\perp,j}/\rho_j B$, where $\rho_j$ is the mass density of species $j$). (Throughout this paper the subscript $j$ will stand for the particle species, $j=i$ for ions and $j=e$ for electrons).This way, on average, the pressures perpendicular ($p_{\perp,j}$) and parallel ($p_{||,j}$) to the local magnetic field must satisfy $p_{\perp,j} > p_{||,j}$. We expect that this anisotropy is ultimately regulated by kinetic micro-instabilities (e.g., ion cyclotron, mirror, firehose, electron whistler, etc), as shown by previous detailed calculations of the kinetic stability of plasmas with anisotropic pressure, PIC simulations \citep[e.g.,][]{GaryEtAl97}, and by solar wind observations \citep{BaleEtAl09}. The kinetic MHD simulations of \cite{SharmaEtAl06, SharmaEtAl07} modeled the presence of these instabilities by setting an upper limit to $|T_{\perp,j}/T_{||,j} - 1|$. This limit on the temperature anisotropy plays a critical role in the evolution of the MRI, making the physics much more MHD-like than it otherwise would have been. Indeed, the saturation of the MRI is qualitatively similar to that in MHD \citep{SharmaEtAl06, SharmaEtAl07}. One significant difference, however, is that the presence of a pressure anisotropy leads to an anisotropic pressure stress that may be as important for angular momentum transport and plasma heating as the magnetic stress.\newline

In this paper, we study the collisionless MRI using first-principle 1D and 2D particle-in-cell (PIC) simulations. We defer more realistic - but also more computationally expensive - 3D calculations to a future paper. In a PIC code, the plasma is represented by a collection of macro-particles that carry charge and mass, and are moved by integration of the Lorentz equations. The electromagnetic fields are evolved by solving Maxwell's equations on a grid, where the current is calculated by adding the contribution of each macro-particle. Given its complete treatment of plasmas, the PIC approach has the ability to capture the whole dynamics of the particles and fields. In particular, 
the MRI, the resulting evolution of the plasma pressure anisotropy, the interaction of particles with small-scale kinetic instabilities, and particle heating and acceleration are all self-consistently captured.\newline 

The reminder of this paper is organized as follows. The basic equations and the simulation setup are explained in \S \ref{sec:baseq} and \S \ref{sec:setup}, respectively.  In \S \ref{sec:injection}, a thorough dispersion relation study is performed based on 1D simulations, which we compare with previous analytic results. We also study the nonlinear magnetic field amplification by the MRI in 1D. In \S \ref{sec:parameters}, we explore the non-linear evolution of the MRI-driven turbulence using more realistic 2D simulations. Special attention is paid to the field saturation, particle heating, pressure anisotropy evolution, and the contribution of the different stress tensor components to angular momentum transport. Finally, in \S \ref{sec:disconclu} we present our conclusions.

\section{Basic Equations}
 \label{sec:baseq}
We carry out our study in the local, small-box approximation, where the size of the simulation box is much smaller than its distance to the center of the disk, $r_0$. The box rotates with the disk at the local (Keplerian) orbital frequency $\omega_0=\omega (r_0)$, so the reference frame is non-inertial. In the rotating frame, Maxwell's equations acquire extra terms \citep{Schiff39}:
\begin{eqnarray}
\nabla \cdot \vec{E} &=& 4\pi \rho_c + \frac{2\vec{\omega}_0 \cdot \vec{B}}{c} - \frac{\vec{v}_0}{c}\cdot \nabla \times \vec{B}, \label{eq:gaussrot}\\ 
\nabla \cdot \vec{B} &=& 0, \label{eq:maggaussrot}\\
\frac{\partial \vec{B}}{\partial t}& = &-c\nabla \times \vec{E}, \textrm{ and} \label{eq:faradrot}\\ 
\frac{\partial \vec{E}}{\partial t} &=& c\nabla \times \vec{B} - 4\pi \vec{J} + \frac{\vec{v}_0}{c} \times \frac{\partial \vec{B}}{\partial t} \nonumber \\
&& -\nabla \times \Big( \vec{v}_0\times\Big( \vec{E} - \frac{\vec{v}_0}{c}\times\vec{B} \Big) \Big),\label{eq:amprot}
\end{eqnarray}
where $\vec{v}_0 = \vec{\omega}_0 \times \vec{r}$ and $c$ is the speed of light.

Our approach is to neglect all of the terms due to the non-inertial reference frame in equations \ref{eq:gaussrot} and \ref{eq:amprot}. We now justify this approximation. In the non-relativistic limit ($v_0 \ll c$), the last two terms on the right hand side of Equation \ref{eq:amprot} are much smaller than $c\nabla \times \vec{B}$. Thus, when $v_0 \ll c$, it is possible to assume that $\vec{J}\approx c\nabla \times \vec{B}/4\pi$. However, because our numerical technique is relativistic (see \S \ref{sec:setup}), we cannot neglect the displacement current in Equation \ref{eq:amprot}, since it is used to evolve $\vec{E}$. We thus choose to integrate Equation \ref{eq:amprot} neglecting the last two terms on the right hand side. Even though this approximation is not expected to affect the MHD-scale dynamics of the plasma, the neglected terms can still formally be larger than the displacement current $\partial \vec{E}/\partial t$, which is not accurately evolved regardless of the $v_0 \ll c$ condition. The effect of this can be seen in Equation \ref{eq:gaussrot}, where the last two terms of the right hand side represent the appearance of extra electric charges in the rotating box. In particular, the term proportional to $\vec{v}_0$ can be much larger than $\nabla \cdot \vec{E}$ if $|\vec{E}|/|\vec{B}| \ll |\vec{v}_0|/c$. This is expected in the case of the small-box approximation, where the typical magnitude of the turbulence velocity ($\sim c|\vec{E}|/|\vec{B}|$) is much smaller than $|\vec{v}_0|$. As a first approach to this problem, we will neglect the existence of theses extra charges, assuming that they do not affect the plasma microphysics. Thus, our simulations will solve the standard Maxwell's equation, with the additional forces due to gravity acting on each particle individually.\newline

In the rotating frame, the particles will experience Lorentz forces, plus the Coriolis and tidal forces; in the case of a Keplerian disk, these are given by the well known expressions: 
\begin{equation}
\frac{d\vec{p}}{dt} = 3m\omega_0^2x \hat{x} - 2\vec{\omega}_0 \times \vec{p} + q(\vec{E} + \frac{u}{c}\times \vec{B}),
\label{eq:cortidal1}
\end{equation}
where $\vec{p}$ and $\vec{u}$ are the particle's momentum and velocity, $m$ and $q$ are its mass and charge, and $x$ corresponds to the radial coordinate.\newline

The expressions for the Coriolis and tidal forces presented in Equation \ref{eq:cortidal1} are valid in the ``cold" limit ($|\vec{u}| \ll |\vec{v_0}|$). Even though in our simulations the particles will reach relativistic velocities, the validity of the cold limit will still hold, but in a fluid sense. This means that, as long as the fluid velocity for each species satisfies $|\vec{u}| \ll |\vec{v_0}|$, the fluid dynamics will be well described by the cold limit expression (Equation \ref{eq:cortidal1}). Also, since in the MRI turbulence $|\vec{u}|$ can be similar to $v_{A}$ ($\equiv B/\sqrt{4\pi \rho_i}$, where the subscript $i$ stands for ions), our non-relativistic, cold limit will be strictly valid when $v_{A} \ll c$. 

\section{Simulation Setup}
\label{sec:setup}
Our simulations are performed using the electromagnetic PIC code TRISTAN-MP \citep{Buneman93, Spitkovsky05} in one and two dimensions. In 2D, the simulation box consists of a rectangle in the $x-z$ plane, where $x$ corresponds to the radial coordinate and $z$ represents the vertical direction of the disk. The azimuthal direction (into the simulation plane) is given by the $y$ axis. The shearing velocity is $\vec{v}=-xs\hat{y}$, where $s$ is the shearing parameter ($\equiv 3\omega_0/2$, in the Keplerian case). \newline 

In standard MHD simulations, shearing periodic boundary conditions are used along the radial ($x$) direction (see, e.g., Hawley et al. 1995). In that case, Galilean transformations of the MHD quantities at the boundaries are made to compensate for their initial velocity difference $|\Delta v|=sL_x$. These shearing periodic boundary conditions can not be self-consistently implemented in PIC simulations. This is because, under a relativistic change of reference frame, the transformation of $\vec{J}$ cannot be obtained by only transforming the velocity of particles. This can introduce charge conservation problems at the box boundaries, even if $|\Delta v|=sL_x \ll c$. We avoid this difficulty by implementing {\it shearing coordinates}, in which the grid moves with the shearing velocity $\vec{v}=-xs\hat{y}$. In this new frame, the net plasma velocity at the boundaries cancels out, allowing the use of periodic boundary conditions. In the shearing coordinate system, the 2D version of Maxwell's equations get modified by the presence of extra terms in Faraday and Ampere's equations. The new equations read 
 \begin{equation}
\frac{\partial \vec{B}(\vec{r},t)}{\partial t} = -\nabla\times\vec{E}(\vec{r},t) - sB_x(\vec{r},t)\hat{y}\textrm{   } \textrm{  } \textrm{    and} 
\label{eq:ind}
\end{equation}
\begin{equation}
\frac{\partial \vec{E}(\vec{r},t)}{\partial t} = \nabla\times\vec{B}(\vec{r},t) -4\pi \vec{J} - sE_x(\vec{r},t)\hat{y}.
\label{eq:amp}
\end{equation}
(see Appendix \ref{sec:shearcoord} for the derivation of equations \ref{eq:ind} and \ref{eq:amp}).\newline

Apart from the modifications to Maxwell's equations, forces on the particles will also transform in the shearing coordinate system:
\begin{equation}
\frac{d\vec{p}}{dt} = 2\omega_0 p_y\hat{x} - \frac{1}{2}\omega_0 p_x\hat{y} +q(\vec{E}+\frac{\vec{u}}{c}\times\vec{B}),
\label{eq:fuerza}
\end{equation} 
where $\vec{p}$ is the particle momentum. We can see that the combination of tidal and coriolis forces are substantially modified, with no dependence on the $x$ coordinate in the shearing frame \footnote{Notice that the combined expression for the coriolis and tidal forces in Equation \ref{eq:fuerza}, and the modified expression for the induction law given by Equation \ref{eq:ind} are equivalent to the 2D versions of equations 14 and 15 of \cite{JohnsonEtAl08}. These equations correspond to the MHD momentum evolution and induction equations, expressed in terms of $\Delta$ {\bf v} $\equiv$ {\bf v} - {\bf v}$_{orb}$, with {\bf v} being the total fluid velocity and {\bf v}$_{orb} = -3x\omega_0/2 \hat{y}$.} (see Appendix \ref{sec:shearcoord} for the derivation of Equation \ref{eq:fuerza}). Thus, our initial set up will consist of a periodic box where, apart from Lorentz forces, particles are pushed by the forces corresponding to the first two terms in the right hand side of Equation \ref{eq:fuerza}, and where the fields are evolved according to Equations \ref{eq:ind} and \ref{eq:amp}.\footnote{Since the modification to Faraday's (Ampere's) equation only affects the evolution of $B_y$ ($E_y$) with an extra term that depends on $B_x$ ($E_x$), we integrate 
$B_y$ ($E_y$) using simple time and space interpolations of $B_x$ ($E_x$). This way, after this modifications are implemented, the numerical algorithm used by TRISTAN-MP continues to be second order accurate in time and space.}\newline

Our simulations are defined by a series of parameters, which set both the physical conditions and numerical resolution of the runs. The physical parameters are the ion to electron mass ratio $m_i/m_e$, the initial magnetic field direction and strength, the orbital frequency $\omega_0$, the initial ion and electron pressures $p_{j}$, and the $x$ and $z$ sizes of the box ($L_x$ and $L_z$). The  initial magnetic field along $\hat{z}$ ($B_{z,0}$) is quantified using the corresponding Alfv\'{e}n velocity of the plasma, $v_{A,0}^z \equiv B_{z,0}/\sqrt{4\pi \rho_i c^2}$). The box size is normalized by $\lambda_0\equiv 2\pi v_{A,0}^z/\omega_0$ (roughly the wavelength of the fastest growing MRI mode in the MHD limit), and the orbital frequency is expressed in terms of the initial ion cyclotron frequency $\omega_{c,i}^z$ ($\equiv |e|B_{z,0}/m_ic$), so our free parameter is the plasma magnetization $\omega_{c,i}^z/\omega_0$. Finally, the initial pressure of the particles is expressed in terms of their initial beta parameter, $\beta_{j}^z$ ($\equiv 8\pi p_{j}/B_{z,0}^2$).\newline

In order to understand our choice of parameters, it is useful to know how they affect the total computing time, $T_{\textrm{comp}}$, of the runs. The computational cost of a simulation is proportional to  $N_{ppc} \times N_{ts} \times N_{gp}$ , where $N_{ppc}$ is the number of particles per cell, while $N_{ts}$ and $N_{gp}$ are the number of times steps and of grid points of the runs, respectively. Thus, it is possible to show that the computing time necessary to run a 2D simulation for a given number of orbital periods, $P_0$ ($\equiv 2\pi/\omega_0$), scales as
\begin{eqnarray} 
T_{\textrm{comp}} & \propto & \Big [(m_i/m_e)^{3/2} (c/v_{A,0}^z)(\omega_{c,i}^z/\omega_0)^3\Big]\times \nonumber\\ 
&&\Big[((c/\omega_{p,e})/\Delta_x)^3(L/\lambda_0)^2 N_{ppc}(\Delta_x/(\Delta_tc))\Big],
\label{eq:scale}
\end{eqnarray}
where $c/\omega_{p,e}$ is the electron inertial length, and $\Delta_x$ and $\Delta_t$ represent the grid spacing and simulation time step, respectively. Equation \ref{eq:scale} shows that the computing time increases for large values of the mass ratio ($m_i/m_e$) and magnetization ($\omega_{c,i}^z/\omega_0$), and for small values of the initial Alfv\'{e}n velocity ($v_{A,0}^z/c$). In addition, there is the increase in computing time due to spatial resolution ($c/\omega_{p,e}/\Delta_x$), box size ($L/\lambda_0$), and particle resolution ($N_{ppc}$). Thus, in general, our simulations will use rather low values for the ion to electron mass ratio, $m_i/m_e$, and magnetization, $\omega_{c,i}^z/\omega_0$, and high values of $v_{A,0}^z/c$. The low mass ratios and magnetizations used in our runs will be significantly far from realistic values. We will assess the dependence of our results on these parameters.
 
\section{1D Simulations: Dispersion Relation Analysis}
\label{sec:injection}
In this Section we study the linear behavior of the collisionless MRI, and compare our results with previous analytical studies. We use 1D simulations, where the $x$ (radial) dimension is reduced to a few cells; this way only wave vectors, $\vec{k}$, pointing along the $z-$axis are resolved. The box length along $z$, $L_z$, is varied in such a way that only one single mode (with $|\vec{k}|=2\pi/L_z$) can grow. The mode is seeded by means of an initial plasma velocity $\vec{v}_{\textrm{seed}}=(v_{A,0}^z/20)\sin(2\pi z/L_z)\hat{x}$, which, by itself, would induce an Alfv\'{e}n wave of linear amplitude in the plasma ($|\delta \vec{B}|/B_{z,0} \sim 1/20$). By measuring the growth rates in each case, we calculate the MRI dispersion relation, which then we compare with previous analytical results. The simulation parameters for each dispersion relation studied are specified in Table \ref{table:1D}. We explore both the weak field regime \citep{KrolikEtAl06, Ferraro07}, and also the high magnetization limit \citep{QuataertEtAl02}. Although our main focus is the highly magnetized case ($\omega_{c,i}^z/\omega_0 \gg 1$), the study of the low magnetization limit will help us understand the parameter regime that minimizes the effect of weak fields, while optimizing the use of computer time. 
\begin{deluxetable}{ccccccc}
\tablecaption{Parameters for the different sets of 1D runs}
\tablehead{ \colhead{Runs} &  \colhead{$\beta_{j}^z$} & \colhead{ $B_{y,0}/B_{z,0}$} & \colhead{$v_{A,0}^z/c$} &
 \colhead{$\omega_{c,i}^z/\omega_0$}  &  \colhead{$m_i/m_e$}
}
\startdata
O1 &  0.05&  0&  1/20 & 11 & 10 \\
O2 &  0.05&  0&  1/20 & -11 & 10 \\
O3 &  0.05&  0&  1/20 & 33 & 10 \\
O4 &  0.05&  0&  1/20 & -33 & 10 \\
O5 &  0.05&  0&  1/20 & 110 & 10 \\
O6 &  0.05&  0&  1/20 & -110 & 10 \\
O7 &  1&  0&  1/20 & 33 & 10 \\
O8 &  1&  0&  1/20 & -33 & 10 \\
O9 &  10&  0&  1/20 & 33 & 10 \\
O10 &  10&  0&  1/20 & -33 & 10 \\
O11 &  0.05&  0&  1/20 & 33 & 1 \\
O12 &  0.05&  0&  1/20 & 33 & 5 \\
O13 &  0.05&  0&  1/20 & 33 & 20 \\
O14 &  0.05&  0&  1/60 & 33 & 10 \\
O15 &  0.05&  1&  1/20 & 33 & 10 \\
O16 &       1&  1&  1/20 & 33 & 10 \\
O17 &       10&  1&  1/20 & 33 & 10 \\
O18 &  1&  0&  1/20 & 220 & 10 \\
O19 &  1&  0&  1/20 & -220 & 10\\
O20 &  1&  0&  1/5 & 33 & 10 \\
O21 &  1&  0&  1 & 33 & 10 \\
O22 &  1&  0&  1/60 & 33 &\textrm{} \textrm{} 10 
\enddata
\tablecomments{We list the beta parameter of ions and electrons $\beta_{j}^z$ (where $j$ stands for ions or electrons), the ratio between the mean $y$ (azimuthal) and $z$ fields $B_{y,0}/B_{z,0}$, the initial Alfv\'{e}n velocity $v_{A,0}^z/c$, the plasma magnetization defined as the ratio of the initial ion cyclotron frequency and the disk rotation frequency $\omega_{c,i}^z/\omega_0$, and the ion to electron mass ratio $m_i/m_e$ ($\beta_{j}^z$, $v_{A,0}^z$, and $\omega_{c,i}^z$ are calculated using only $B_{z,0}$). The space and particle resolutions in all of our 1D simulations are given by $c/\omega_{p,e}/\Delta_x=10$ and $N_{ppc}=15$.}
\label{table:1D}
\end{deluxetable}

\subsection{Low Magnetization Regime}
\label{sec:mhdlimit}
\begin{figure}
\begin{center}
\centering\includegraphics[width=8cm]{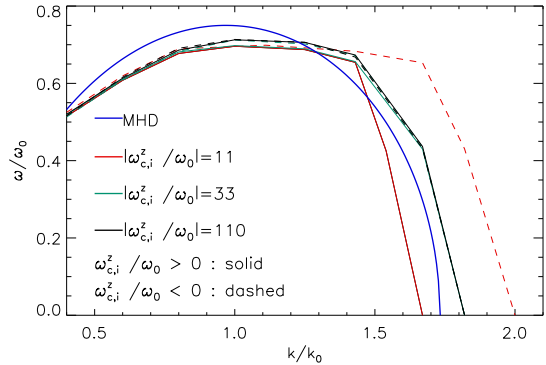}
\caption{Dispersion relations for 1D simulations O1-O6 from Table \ref{table:1D}, which use the same parameters ($\beta_{j}^z=0.05$, $v_{A,0}^z/c=1/20$, $m_i/m_e=10$, and $B_{y,0}=0$), except for the initial magnetization of the plasma, which is defined by the ratio of the initial ion cyclotron frequency to the disk orbital frequency $\omega_{c,i}^z/\omega_0$. The red, green, and black lines are for $|\omega_{c,i}^z/\omega_0|= 11$, 33, and 110; solid and dashed lines show $\omega_{c,i}^z/\omega_0 > 0$ and $< 0$, respectively. The results converge at $|\omega_{c,i}^z/\omega_0|= 33$ to a dispersion relation reasonably in agreement with the analytical MHD prediction \citep{BalbusEtAl91}, shown with the blue line. The growth rate $\omega$ is normalized in terms of the orbital frequency $\omega_0$, and the wavenumber $k$ is normalized in terms of $k_0$ ($\equiv \omega_0/v_{A,0}^z$). These numerical results are consistent with the analytical dispersion relation calculation of \cite{KrolikEtAl06}.}
\label{fig:magnetconverg}
\end{center}
\end{figure}

\begin{figure}
\begin{center}
\centering\includegraphics[width=8cm]{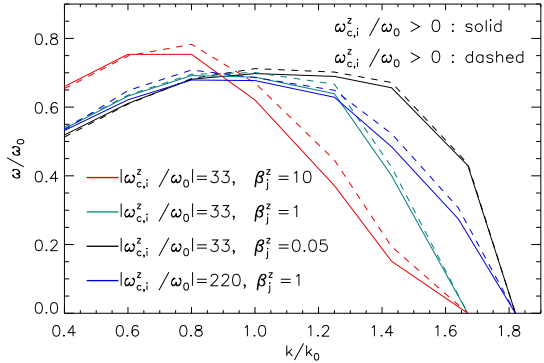}
\caption{The black, green, and red lines show dispersion relations for 1D simulations with the same initial plasma magnetization $|\omega_{c,i}^z/\omega_0|=33$, but with different plasma betas: $\beta_{j}^z=0.05$ (O3 and O4), 1 (O7 and O8), and 10 (O9 and O10); solid and dashed lines show $\omega_{c,i}^z/\omega_0 > 0$ and $< 0$, respectively. As $\beta_j^z$ increases, the unstable MRI modes shift to larger wavelengths, with the maximum growth rate increasing. This is a consequence of the ion Larmor radius increasing with increasing $\beta_j^z$. There is no substantial difference between $\omega_{c,i}^z/\omega_0 > 0$ and $<0$. The blue lines show runs like O7 and O8, but with 6 times larger magnetization ($|\omega_{c,i}^z/\omega_0|=220$; runs O18 and O19). The migration to longer wavelengths seen in runs O7 and O8 is reduced, and the dispersion relations approach the $\beta_{j}=0.05$ cases. This is consistent with the fact that relatively large values of $\beta_{j}$ do not produce significant finite Larmor radius (FLR) effects if the magnetization is large enough.}
\label{fig:betadepen}
\end{center}
\end{figure}

The low magnetization regime has been explored analytically both in the cold limit \citep{KrolikEtAl06}, and in the finite temperature case \citep{Ferraro07}. We will investigate the effect of $\omega_{c,i}^z/\omega_0$ in the cold case first. We measure the MRI dispersion relation for $\beta_{j}^z=0.05$ using the magnetizations $\omega_{c,i}^z/\omega_0 = 11$, -11, 33, -33, 110, and -110 (corresponding to simulations O1-O6 in Table \ref{table:1D}), which are presented in Figure \ref{fig:magnetconverg}. The red lines correspond to $|\omega_{c,i}^z/\omega_0| = 11$ with the solid (dashed) line depicting the $\omega_{c,i}^z/\omega_0 > 0$ ($<0$) case. Although neither the wave number of the fastest growing mode nor the corresponding growth rate varies between these two cases, the range of unstable wave numbers extends to larger values when $\omega_{c,i}^z/\omega_0=-11$. The green and black lines show results for $|\omega_{c,i}^z/\omega_0| = 33$ and 110, respectively. There is practically no difference between these two magnetizations, and the sign of $\omega_{c,i}^z/\omega_0$ no longer plays a role. This shows that, when $|\omega_{c,i}^z/\omega_0| = 33$, our simulations have already converged to a dispersion relation reasonably in agreement with the analytical MHD result \citep{BalbusEtAl91}. The way the dispersion relation depends on the sign and magnitude of  $\omega_{c,i}^z/\omega_0$ shows that at low magnetization the coupling between the particles' gyromotion and their epicycle motion can modify the MRI dynamics significantly. These results are consistent with the MRI dispersion relations at low magnetization (zero temperature) presented in Figures 1 and 2 of  \cite{KrolikEtAl06}.\newline
We have also studied the effect of finite particle temperatures (and, thus, of finite Larmor radius) by re-running simulations O3 and O4 using $\beta_{j}^z=$1 (runs O7 and O8) and 10 (runs O9 and O10). The results are presented in Figure \ref{fig:betadepen}.  The black lines show the `cold' ($\beta_{j}^z=0.05$) case of Figure \ref{fig:magnetconverg}, and the green and red lines show the $\beta_{i,e}^z=1$ and 10 results. The solid and dashed lines show $\omega_{c,i}^z/\omega_0 > 0$ and $<0$, respectively. Figure \ref{fig:betadepen} shows that for larger initial plasma pressure, the range of unstable MRI modes shifts to larger wavelengths. The maximum value of the growth rate also increases for larger initial pressure. In addition, no substantial difference is observed between the $\omega_{c,i}^z/\omega_0 > 0$ and $<0$ cases. This result can be compared with the analytical treatment of \cite{Ferraro07}, where the effect of finite Larmor radii (FLR) of the ions was included. Ferraro's analytical prediction is consistent with our numerical calculation only for $\omega_{c,i}^z/\omega_0 > 0$ (see his Figure 1). For $\omega_{c,i}^z/\omega_0 < 0$, his prediction is that larger temperatures increase the upper limit of the unstable wave numbers and reduce the maximum growth rate. We do not find this dependence on the sign of $B_{z,0}$. This may be because our simulations use electrons in energy equipartition with the ions, in contrast with the perfectly cold electrons used by \cite{Ferraro07}. In any case, if the dependence on $\beta_{j}^z$ in Figure \ref{fig:betadepen} is caused by FLR effects of the kind predicted by \cite{Ferraro07}, then increasing the plasma magnetization for fixed $\beta_{j}^z$ should cause our results to approach the limit in which FLR effects are negligible. We tested this by taking runs O7 and O8 (with $\beta_{i,e}^z=1$) and increasing their magnetization by a factor of 6 to $|\omega_{c,i}^z/\omega_0|=220$ (while keeping the same $\beta_{j}^z$). The corresponding dispersion relations are shown by the blue lines in Figure \ref{fig:betadepen}. Increasing the magnetization indeed increases the range of unstable wave numbers, with the results approaching the cold plasma (zero FLR effects) results. Thus, for the magnetizations utilized in this paper, we expect FLR effects to be present for finite values for $\beta_{j}$. In the astrophysical regimes of interest, however, FLR effects are expected to be negligible.    

\subsection{High Magnetization Regime}
\label{sec:kmhdlimit}
\begin{figure}
\begin{center}
\centering\includegraphics[width=8cm]{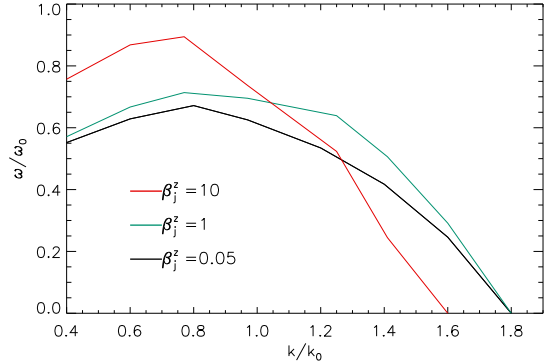}
\caption{Dispersion relation calculations for runs with $B_{y,0}=B_{z,0}$ but different values of $\beta_{j}^z=0.05, 1$, and 10 (black, green, and red lines corresponding to runs O15, O16, and O17, respectively). The growth of longer wavelength modes is favored at higher pressures, with the maximum growth rate for the $\beta_{j}^z=10$ case being significantly larger than for $B_{y,0}=0$ (shown by the red lines in Figure \ref{fig:betadepen}). This is consistent with linear analytic calculations \citep{QuataertEtAl02}.}
\label{fig:betadepenbphi1}
\end{center}
\end{figure}

The linear behavior of the MRI has been studied analytically in the high magnetization regime by \cite{QuataertEtAl02; SharmaEtAl03} using the kinetic MHD approach. In this approach, it is assumed that $\omega_{c,i}^z/\omega_0 \to \infty$ and also that FLR effects are unimportant, i.e. the ion gyroradius is much smaller than $\lambda_0$. One of the main differences with respect to standard MHD is the increase in both the growth rate and the wavelength of the fastest growing mode, which happens for large $\beta_{j}^z$ in the presence of a significant toroidal magnetic field. We tested this result for simulations with $\beta_{j}^z =0.05, 1,$ and 10, with $B_{y,0}/B_{z,0}=$1 (runs O15, O16, and O17, respectively). The corresponding dispersion relations are shown in Figure \ref{fig:betadepenbphi1}, where the black, green, and red lines depict the cases with $\beta_{j}^z =0.05, 1,$ and 10. The tendency to favor the growth of longer wavelengths and for larger maximum growth rates for larger $\beta_j^z$ is clearly seen. Notice that the maximum growth rate for the $\beta_{j}^z=10$, $B_{y,0}=B_{z,0}$ case is significantly larger than for the analogous case with $B_{y,0}=0$ (shown in Figure \ref{fig:betadepen}). These dispersion relations are in reasonable agreement with their analytical counterpart shown in the right panel of Figure 4  of \cite{QuataertEtAl02}.\newline

\subsection{Dependence on other Parameters}
\label{sec:massratio}

\begin{figure}
\begin{center}
\centering\includegraphics[width=8cm]{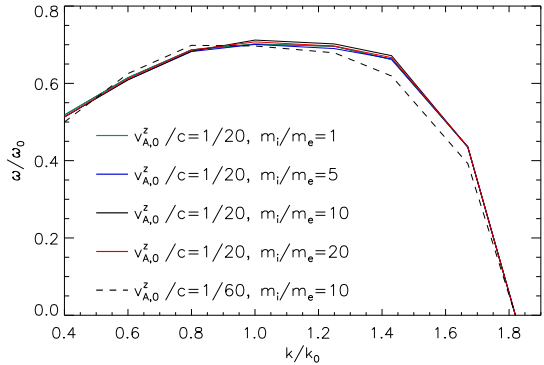}
\caption{A comparison of numerical dispersion relations for different  mass ratios $m_i/m_e=1$ (green; run O11), $5$ (blue; run O12), 10 (black; run O3), and 20 (red; run O13). The value of $m_i/m_e$ does not change the linear growth rate, even for $m_i/m_e=1$. Also, a test of the effect of $v_{A,0}^z/c$ is shown by the dashed-black line, which corresponds to $v_{A,0}^z/c=1/60$ (run O14). No substantial difference is observed when comparing with the $v_{A,0}^z/c=1/20$ cases.}
\label{fig:massratdepen}
\end{center}
\end{figure}

Given the low mass ratios $m_i/m_e$ that we use, it is important to check that our results are not affected by this parameter. Thus, we carried out simulations analogous to run O3 but using different values of $m_i/m_e$. The dispersion relations are shown in Figure \ref{fig:massratdepen} for $m_i/m_e=1, 5$, 10 and 20 (green, blue, black, and red lines, respectively). Figure \ref{fig:massratdepen} shows that $m_i/m_e$ does not play any significant role in the linear dispersion relation of the MRI, which does not change between $m_i/m_e=1$ and 20. Finally, in the same figure, a test of the effect of $v_{A,0}^z/c$ is shown by the dashed-black line, which shows the dispersion relation for $v_{A,0}^z/c=1/60$ (run O14). There is no substantial difference relative to the $v_{A,0}^z/c=1/20$ cases.\newline

\subsection{MRI Saturation in 1D}
\label{sec:saturation}
Most of our non-linear MRI analysis will be done using 2D simulations. In this section, however, we determine a saturation criterion for the magnetic amplification in 1D. Although far from realistic, this will help us better understand and interpret the 2D results in \S \ref{sec:parameters}. 
\begin{figure}
\begin{center}
\centering\includegraphics[width=8cm]{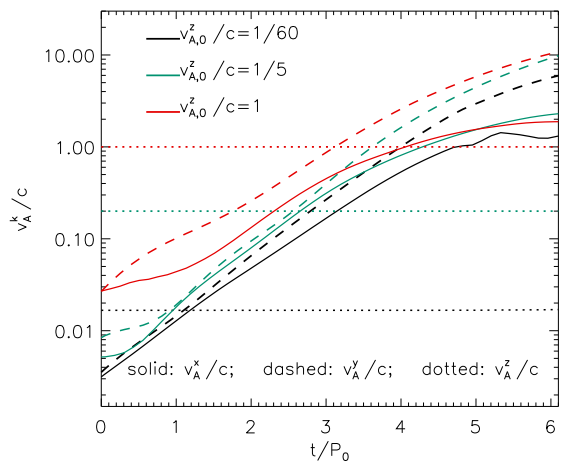}
\caption{The Alfv\'{e}n velocities $v_A^k/c$ calculated with the $k$ component of the magnetic field ($k=x,y,$ and $z$) in three 1D simulations (O22, O20, and O21). The runs have the same box-size $L_z=1.25 \lambda_0$ ($\lambda_0 \equiv 2\pi v_{A,0}^z/\omega_0$) and the same parameters, except for their initial $v_{A,0}^z/c$. The black, green and red lines correspond to $v_{A,0}^z/c= 1/60, 1/5$, and 1 (runs O7, O20, and O21 of Table \ref{table:1D}, respectively). All three calculations have roughly the same saturation, with $v_A^x \sim c$ and $v_A^y \sim 10c$. Similar results are obtained using a larger box ($L_z=5 \lambda_0$), and at different $m_i/m_e$ and magnetizations.}
\label{fig:saturation}
\end{center}
\end{figure}
Figure \ref{fig:saturation} shows the magnetic energy evolution for three 1D simulations with ``box" sizes $L_z=1.25 \lambda_0$ and with similar parameters except for their initial $v_{A,0}^z/c$. The black, green and red lines correspond to $v_{A,0}^z/c= 1/60, 1/5$, and 1 (runs O22, O20, and O21 of Table \ref{table:1D}),  and the solid and dashed lines show the evolution of their radial and azimuthal  magnetic energies, $B_x^2/B_0^2$ and $B_y^2/B_0^2$, respectively.  The maximum amplification of $B_x$ in all three cases satisfies $v_{A}^x \sim c$ (where $v_{A}^x$ corresponds to the Alfv\'{e}n velocity calculated only with the $x$ component of the magnetic field). After the saturation of the radial field, $B_y$ continues to grow, but at a significantly lower rate. This result appears to be independent of the size of the box (it was also tested for $L_z=5 \lambda_0$), and other parameters like $m_i/m_e$, $\omega_{c,i}^z/\omega_0$, and $\beta_{j}$.  Note also that for $v_{A,0}^z/c= 1$ the linear growth rate is reduced, leading to a suppressed exponential growth of $B_x$.\newline 
It is important to emphasize that, by assumption, our treatment of the MRI is valid only in the non-relativistic regime, i.e., when $v_A \ll c$. This regime is the most interesting since $v_A \sim c$ corresponds to a magnetic field energy close to the rest mass energy of the particles, which should be precluded by energy conservation, except very close to a black hole event horizon. Thus, this saturation criterion implies that, at least in 1D, there is no mechanism stopping the growth of the field.\newline
  
In the next section we study how this 1D evolution is modified by 2D effects. We also analyze the sources of angular momentum transport in detail, and study the interplay between the non-linear MRI turbulence and particle heating. 

\section{Two-dimensional Simulations}
\label{sec:parameters}
Our 2D analysis is organized in four parts. First, \S \ref{nonlinear} describes the overall non-linear behavior of the MRI turbulence, paying special attention to its saturation. \S \ref{zeroflux} briefly discusses how the non-linear evolution is modified in the zero magnetic flux case. \S \ref{stresses} analyses angular momentum transport, considering the contribution of an anisotropic pressure stress. Finally, in \S \ref{heating} we discuss particle heating and identify the different processes that contribute to it.\newline
Our analysis is based on a series of simulations listed in Table \ref{table:2D}. The initial physical conditions of the runs are defined by: the beta of ions and electrons $\beta_{j}^z$, the magnetic field along $\hat{y}$, $B_{y,0}$, the field along $\hat{z}$, $B_{z,0}$ (quantified via $v_{A,0}^z/c$), the plasma magnetization $\omega_{c,i}^z/\omega_0$, and the ion to electron mass ratio, $m_i/m_e$ ($v_{A,0}^z/c$, $\beta_{j}^z$, and $\omega_{c,i}^z$ are calculated only considering $B_{z,0}$). The remaining parameters determine the numerical resolution of the runs. These are defined by: the box dimensions ($L_x\times L_z)/\lambda_0^2$ (where, as before, $\lambda_0=2\pi v_{A,0}^z/\omega_0$), and the space, time, and particle resolutions. The space and time resolutions are set by the number of grid points per electron skin depth, $c/\omega_{p,e}/\Delta_x$, while the particle resolution is defined by the number of particles per cell, $N_{ppc}$. 

\begin{deluxetable*}{cccccccccc}
\tablecaption{Parameters of the 2D runs}
\tablehead{ \colhead{Runs} &  \colhead{$\beta_{j}^z$} & \colhead{ $B_{y,0}/B_{z,0}$} & \colhead{$v_{A,0}^z/c$} &
 \colhead{$\omega_{c,i}^z/\omega_0$}  &  \colhead{$m_i/m_e$} & \colhead{$L_x\times L_z /\lambda_0^2$} & \colhead{$c/\omega_{p,e}/\Delta_x$} & \colhead{$N_{ppc}$} & \colhead{zero flux?}
 }
\startdata
T1 &  1&  0&  1/20 & 11 & 2 & $8\times8$ & 7 & 3 &no\\
T2 &  1& 0& 1/20& 11 & 5 & $4\times4$ & 7 & 3 &no\\
T3 &  1&  0&  1/60 & 11 & 2 & $5\times5$ & 7 & 3 &no\\
T4 &  40&  0&  1/20 & 22 & 2 & $4\times4$ & 7 & 3 &no\\
T5 &  1&  1&  1/20 & 11 & 2 & $8\times8$ & 7 & 3 &no\\
T6 &  1&  0&  1/20 & 22 & 2 & $4\times4$ & 7 & 3 &no\\
T7 &  1&  0&  1/120 & 11 & 2 & $5\times5$ & 7 & 3 &no\\
T8 &  10&  0&  1/20 & 11 & 2 & $8\times8$ & 7 & 3 &no\\
T9 &  1&  0&  1/20 & 11 & 2 & $2\times2$ & 10 & 6 &no\\
T10 &  1&  0&  1/20 & 11 & 2 & $4\times4$ & 7 & 3 &no\\
T11 &  10&  0&  1/20 & 11 & 2 & $2\times2$ & 14 & 3 &no\\
T12 &  1&  0&  1/60 & 11 & 2 & $8\times8$ & 7 & 3 &yes\\
T13 &  1&  0&  1/20 & 11 & 2 & $8\times8$ & 7 & 3 &yes\\
T14 &  1&  0&  1/20 & 11 & 2 & $16\times8$ & 7 & 3 &\textrm{ yes}
\enddata
\tablecomments{A list of 2D simulations, defined by the initial beta parameter of ions and electrons $\beta_{j}^z$, the ratio between the mean $y$ (azimuthal) and $z$ (vertical) fields, $B_{y,0}/B_{z,0}$, the initial Alfv\'{e}n velocity, $v_{A,0}^z/c$, the plasma magnetization, $\omega_{c,i}^z/\omega_0$ (the ratio between the initial ion cyclotron frequency and the rotation frequency of the disk), and the ion to electron mass ratio, $m_i/m_e$ (the $z$ superscript indicates that $\beta_{j}^z$, $v_{A,0}^z/c$, and $\omega_{c,i}^z$ are defined by the $z$-component of $\vec{B}$). The numerical resolution of the runs is defined by the box size $L_x/\lambda_0$ and $L_z/\lambda_0$ (where $\lambda_0 \equiv 2\pi v_{A,0}^z/\omega_0$), and the space and particle resolutions, which are determined by the number of grid points per electron skin depth $c/\omega_{p,e}/\Delta_x$ and the number of particles per cell $N_{ppc}$, respectively.}
\label{table:2D}
\end{deluxetable*}

\subsection{MRI Turbulence Evolution}
\label{nonlinear}
\begin{figure}
\begin{center}
\centering\includegraphics[width=8cm]{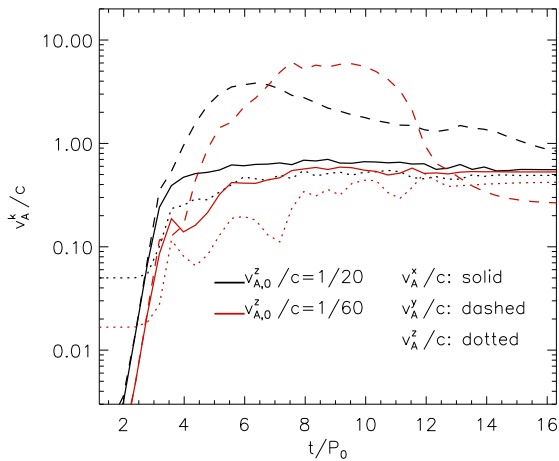}
\caption{Evolution of the three components of the magnetic field for 2D simulations T1 ($v_{A,0}^z/c=1/20$; black lines) and T5 ($v_{A,0}^z/c=1/60$; red lines). Magnetic field values are expressed in terms of the Alfv\'{e}n velocities $v_A^k$ ($\equiv B_k/\sqrt{4\pi \rho}$, with $k=x, y,$ and $z$). The solid, dashed, and dotted lines represent the $x$, $y$, and $z$ components, respectively. As in 1D, the magnetic field saturates at $v_A^{x,z}/c \sim 1$ and $v_A^{y}/c \sim 10$.}
\label{fig:magevocompare}
\end{center}
\end{figure}

\begin{figure*}
\begin{center}
\centering\includegraphics[width=17cm]{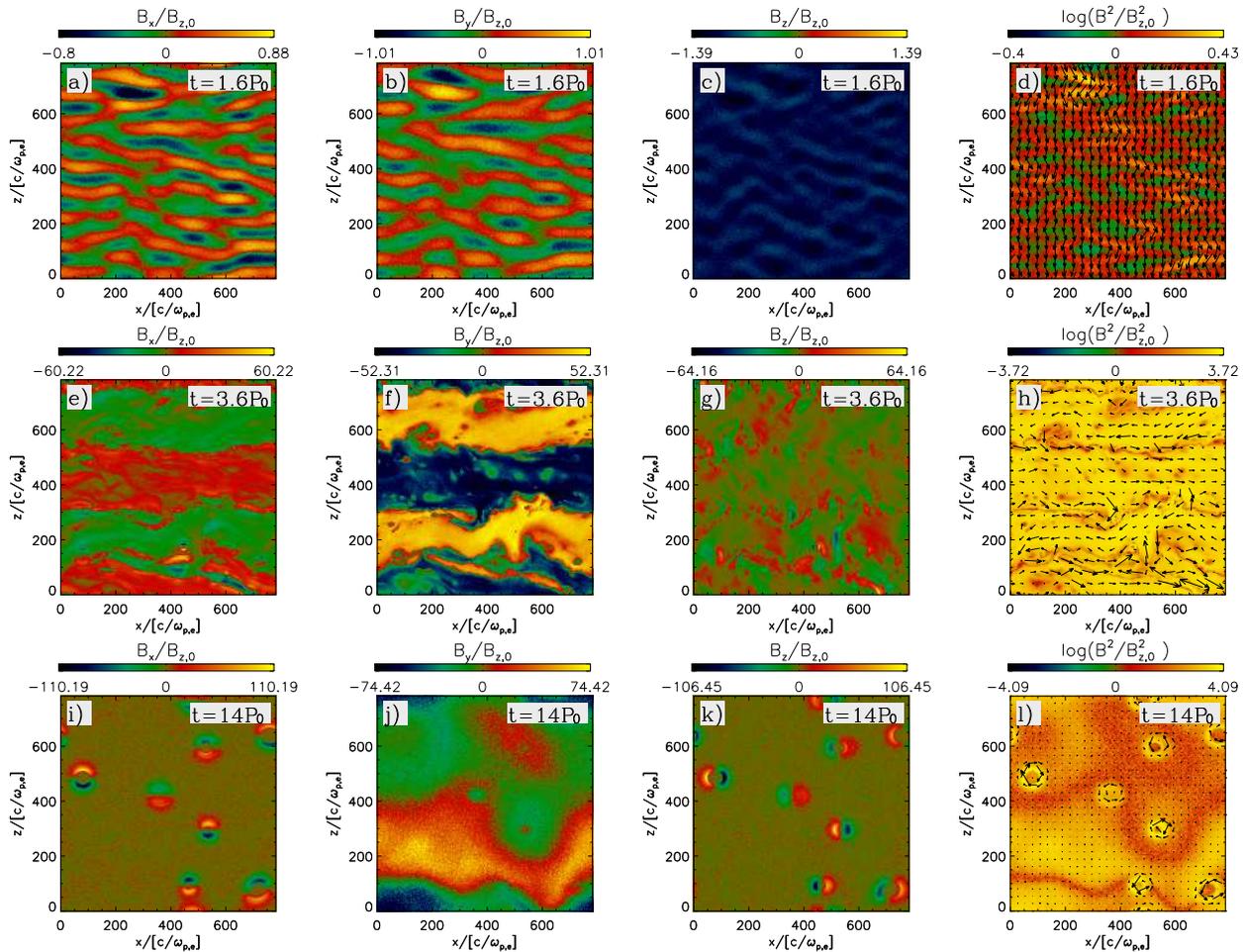}
\caption{The three components of the magnetic field $B_x$, $B_y$, and $B_z$ and its energy normalized in terms of the initial field $B_{z,0}$, for run T1 at three different times. The arrows in the $\log(B^2/B_{z,0}^2)$ plots show the projection of the magnetic field direction on the $x-z$ plane. At $t=1.6P_0$  (top row)the MRI is in the mildly non-linear regime, with the magnetic fluctuations dominated by the fastest growing linear MRI channel mode. At $t=3.6P_0$ (middle row), the growth has migrated to longer wavelengths, with the dissipation of the short wavelength modes dominated by magnetic reconnection. Finally, at $t=14P_0$ (bottom row), the turbulence is in a quiescent state, with no MRI modes and magnetic field concentrated in loops. At that point, growing MRI modes have wavelengths larger than the box size.}
\label{fig:mri2dBxByBzB2}
\end{center}
\end{figure*}

The non-linear MRI evolution is characterized by an initial exponential growth of the field (until $|\vec{B}|/B_{z,0} \sim 5$), followed by a significant decrease in the growth rate. This can be seen in Figure \ref{fig:magevocompare}, which shows the evolution of the three magnetic energy components for simulations T1 (black) and T3 (red)  of Table \ref{table:2D} quantified using their Alfv\'{e}n velocities ($v_A^x/c$, $v_A^y/c$, and $v_A^z/c$, represented by solid, dashed, and dotted lines, respectively). 
In both cases, while $v_A^y/c$ continues to grow exponentially (although at a slower rate), the $x$ and $z$ components appear to enter a linear growth regime and saturate at amplitudes about one order of magnitude smaller than the azimuthal field. In both cases, the magnetic growth proceeds until $v_{A}^x \approx 2v_{A}^z \approx c$ and $v_{A}^y \approx 10c$, which is qualitatively consistent with the 1D saturation criterion found in \S \ref{sec:saturation}. Indeed, there are only two important differences relative to the 1D case. The first one is the growth of $B_z$, which can not occur in the 1D case ($\nabla \times \vec{E}$ can not have a component along $\hat{z}$, given that $\partial_x=\partial_y=0$). The other difference is that, regardless of the initial $v_{A,0}^z$, the exponential field growth stops at $|\vec{B}|/B_{z,0} \approx 5$. This can be explained by the role of magnetic reconnection in dissipating the field energy in the non-linear regime of our 2D runs. Figure \ref{fig:mri2dBxByBzB2} shows the 2D structure of the magnetic field for run T1, with the three field components along with the magnetic energy plotted at $t=1.6$, 3.6 and $14P_0$.
At $t=1.6 P_0$ the instability is in the mildly non-linear regime ($|\vec{B}|/B_{z,0} \sim 1$). At that point, magnetic amplification is dominated by the fastest growing MRI mode with a wavelength $\lambda \approx \lambda_0$. At $t=3.6P_0$, on the other hand, the field has been amplified to  $|\vec{B}|/B_{z,0} \sim 60$ and the dominant wavelength has grown to $\lambda \approx 4\lambda_0$. This migration to longer wavelengths occurs along with the reconnection of magnetic field lines associated with the small wavelength modes as they become non-linear. Magnetic reconnection takes place in thin current sheets, where $B_x$ and $B_y$ switch sign. Along with the onset of reconnection, there is the appearance of loop-like structures (see, for instance, the one at $(x,z) \approx (200,200)c/\omega_{p,e}$ in Figures \ref{fig:mri2dBxByBzB2}$e$-\ref{fig:mri2dBxByBzB2}$h$). The magnetic loops also correspond to regions of high plasma density, as can be seen from Figure \ref{fig:mri2dBxdens}, which shows the evolution of the plasma density, $\rho/\rho_0$, and $B_x/B_0$ for run T1. The formation of loops is much clearer at $t=14P_0$. At that stage, they appear as overdense regions ($\rho/\rho_0 \sim 10$), in pressure equilibrium with the surrounding magnetic field. \newline

As $B_y$ gradually decays, the magnetic energy corresponding to $B_x$ and $B_z$ stays rather constant. This energy is contained primarily in the magnetic loops, as can be seen from the field plots at $t=14P_0$ in Figures \ref{fig:mri2dBxByBzB2}$i$-\ref{fig:mri2dBxByBzB2}$l$. At this point, the fields are in a quiescent state, with the loops experiencing almost no evolution (as can also be inferred from the smooth magnetic energy evolution after $t \gtrsim 13$ in Figure \ref{fig:magevocompare}). This implies that, at late times, the growth of new linear MRI modes is dramatically suppressed. This behavior can be explained by the migration of the growing modes to large wavelengths due to finite Larmor radius (FLR) effects at large $\beta_{j}^z$ (see \S \ref{sec:mhdlimit}). Indeed, for the relatively low magnetization $\omega_{c,i}^z/\omega_0=11$ of simulation T1, the observed increase in temperature corresponds to $\beta_{j}^z\approx 1000$ (when only the initial $B_{z,0}$ is considered), which makes the particle's gyroradii outside of the loops larger than $\lambda_0$ (after $B_y$ has been significantly dissipated). Thus, FLR effects should significantly increase the wavelength of the unstable modes, presumably to length-scales larger than the box size. In a realistic astrophysical scenario, however, the MRI will not be suppressed by FLR effects. Indeed, $\omega_{c,i}^z/\omega_0$ is typically many orders of magnitude larger than the value used in our simulations, which would make FLR effects negligible \citep{Ferraro07}. \newline

Also, in all of our simulations the quiescent state happens after the MRI has reached the saturation condition $v_{A}^x \approx 2v_{A}^z \approx c$ and $v_{A}^y \approx 10c$. However, we expect the MRI to saturate before reaching this condition in a more realistic 3D problem. Indeed, we believe that the 2D geometry of our simulations favors both the different evolution of $B_y$ (compared with $B_x$ and $B_z$) and the unrestricted field growth in the non-relativistic regime ($v_A \ll c$). 
Loop formation makes reconnection a 2D phenomenon, which is favored if these structures are well resolved by the simulation. In our 2D runs, this is the case only for reconnection of field lines laying mainly on the $x-z$ plane. 
\begin{figure}
\begin{center}
\centering\includegraphics[width=8cm]{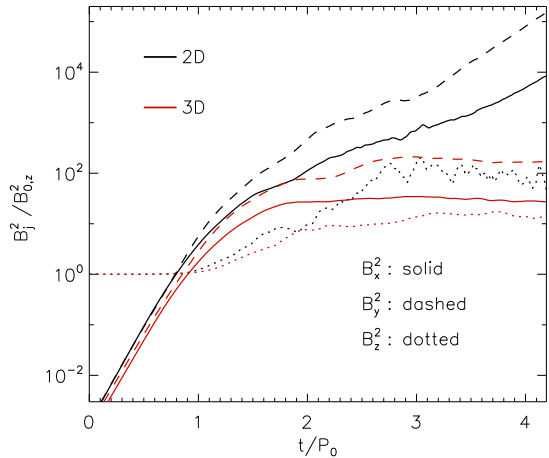}
\caption{Comparison of magnetic energy evolution in 2D (black) and 3D (red) runs that use the kinetic MHD model of \cite{SharmaEtAl06}. These fluid simulations are analogous to simulation T1, i.e., they use the same initial $\beta$ and box size $L_{x,z}/\lambda_0$ (with the 3D case having $L_y=L_{x,z}$). The $x$ (radial), $y$ (azimuthal), and $z$ (vertical) components of the magnetic energy are shown with solid, dashed, and dotted lines, respectively. In contrast to the 3D case, the 2D case does not saturate, qualitatively reproducing the lack of saturation in our 2D PIC simulations when $v_A < c$. Also, in the 2D case the magnetic energy is dominated by the azimuthal field component (black-dashed line), which is also in agreement with our PIC results. This suggests that the 2D geometry of our runs plays a crucial role in precluding field saturation in the $v_A < c$ regime.}
\label{fig:t1mhd2d3d}
\end{center}
\end{figure} 
The effect of the 2D geometry can be seen in Figure \ref{fig:t1mhd2d3d}, which shows the magnetic energy evolution of 2D and 3D kinetic MHD versions of run T1 \citep[using the same modified version of the ZEUS code as in ][]{SharmaEtAl06}. We see that the magnetic field energy evolves fundamentally differently in the 2D and 3D runs. In the 2D case, the magnetic energy growth does not saturate, with the $B_y$ contribution dominating in the non-linear regime (like in our runs); while in the 3D case the field saturates at $|\delta \vec{B}/B_{z,0}| \sim 10$ with $B_y$ contributing nearly the same energy as the other two field components.
\begin{figure}
\begin{center}
\centering\includegraphics[width=8.5cm]{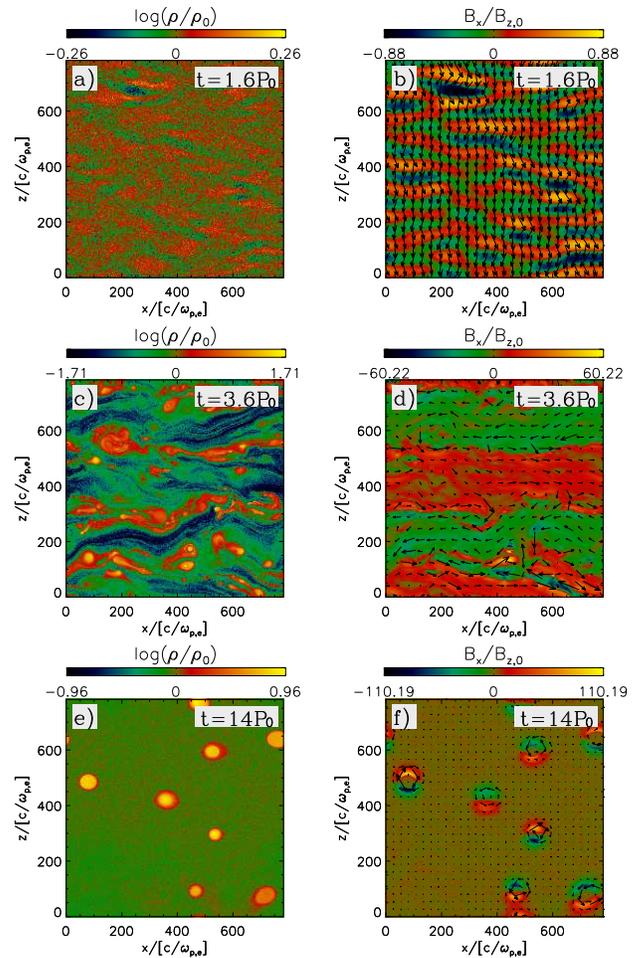}
\caption{The plasma density, $\rho/\rho_0$, and $B_x$, for our fiducial 2D simulation T1 at three different times. The arrows in the $B_x$ plot represent the magnetic field direction on the $x-z$ plane. At $t=1.6P_0$ (top row) the instability is in the mildly non-linear regime. At $t=3.6P_0$ (middle row), the short wavelength modes have dissipated by magnetic reconnection and growth has migrated to longer wavelengths. Finally, at $t=14P_0$ (bottom row), the turbulence has died away and growing MRI modes have wavelengths larger than the box size. In this quiescent state, both the magnetic field and plasma are concentrated in loops created by reconnection.}
\label{fig:mri2dBxdens}
\end{center}
\end{figure}  
\begin{figure}
\begin{center}
\centering\includegraphics[width=8.5cm]{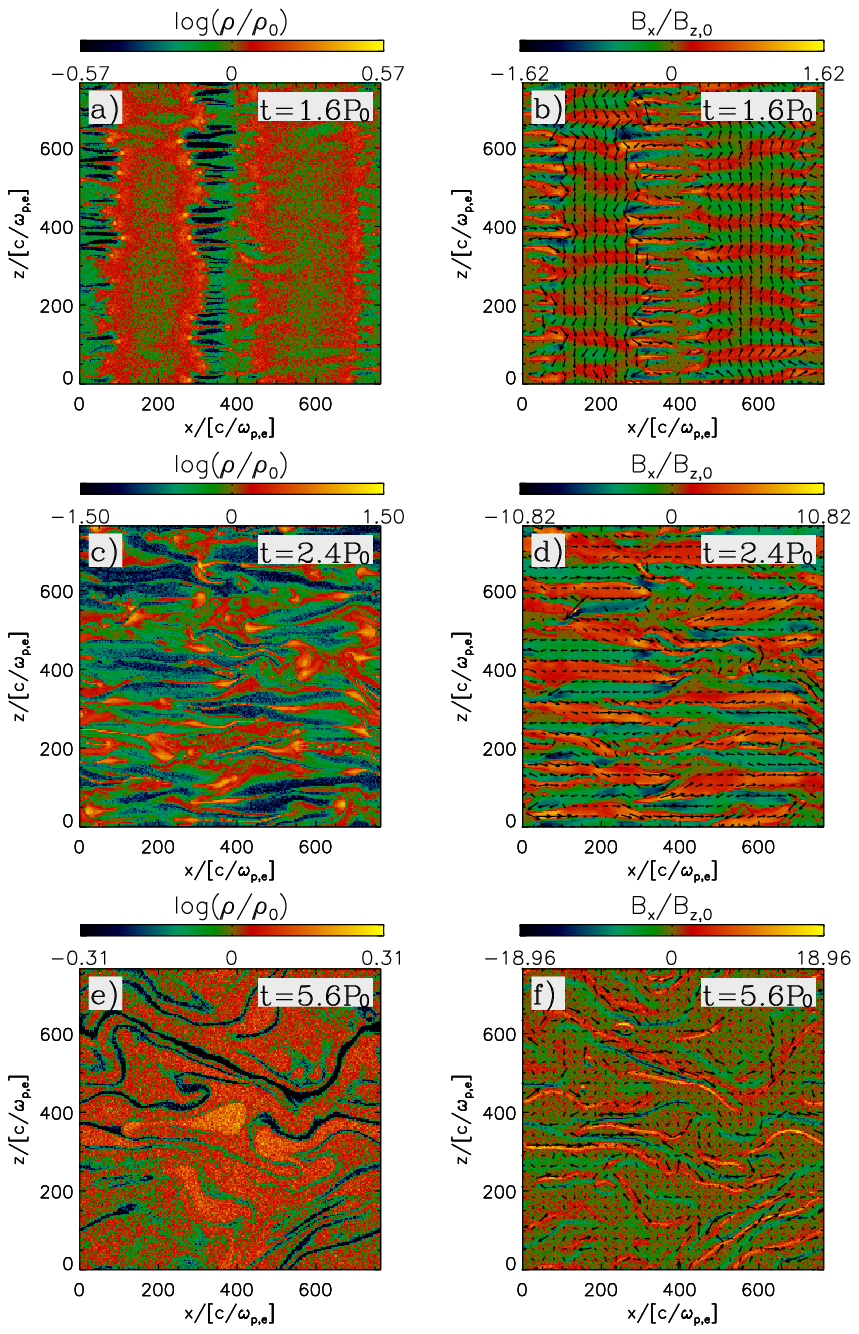}
\caption{The plasma density, $\rho/\rho_0$, and $B_x$, for 2D run T12 at three different times, showing the overall MRI evolution in the case of zero net $B_z$ flux. The first, second, and third rows show the linear, non-linear, and post-saturation states at $t=1.6P_0$, $2.4P_0$, and $5.6P_0$, respectively. With zero-net flux reconnection is much more vigorous and leads to saturation of the MRI prior to $v_A \sim c$ (see Figure \ref{fig:magevocomparenoflux}). Magnetic loops are also much less prominent.}
\label{fig:mri2dBxdensnoflux}
\end{center}
\end{figure} 

\subsection{The zero net flux case}
\label{zeroflux}
The MRI evolution presented above can be substantially modified if the net magnetic flux along $\hat{z}$ is zero. Figure \ref{fig:mri2dBxdensnoflux} shows the density and $B_x$ of run T12, which is analogous to run T3 but with $\vec{B}_0 = -\sin(x/L_x)B_{z,0} \hat{z}$. In the zero net flux case the MRI initially grows faster in the low field regions. This is due to finite Larmor radius (FLR) effects, as explained in \S \ref{sec:mhdlimit}. When $|B_0|$ is small, the beta of the plasma is large, which increases the growth rate of the fastest growing mode (see Figure \ref{fig:betadepen}). This effect also appears to be stronger when $\omega_{c,i}^z/\omega_0 < 0$, which is consistent with the slightly different growth rates seen in Figure \ref{fig:betadepen} for different signs of $\omega_{c,i}^z/\omega_0$. This FLR effect is not expected in realistic astrophysical settings. \newline
\begin{figure}
\begin{center}
\centering\includegraphics[width=8cm]{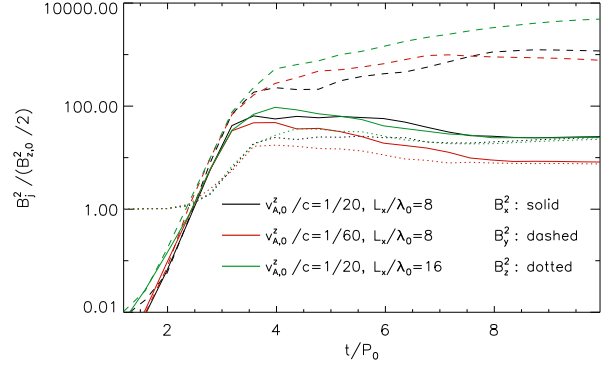}
\caption{The evolution of the three magnetic energy components for simulations T12, T13, and T14, with no vertical flux. The maximum magnetic field amplification ($B_{max}/B_{z,0}$) in runs T12 ($v_{A,0}^z/c=1/60$; black line) and T13 ($v_{A,0}^z/c=1/20$; red line) appear to be almost the same, despite their different initial Alfv\'{e}n velocity. This is fundamentally different from the finite $B_z$ flux cases, where the lower the initial $v_{A,0}^z$, the larger the amplification factor (so that at saturation $v_{A,0}^x \sim v_{A,0}^z \sim c$, and $v_{A,0}^y \sim 10c$). The amplification seems to increase in the case of run T14 ($v_{A,0}^z/c=1/20$; green line), which has a $L_x/\lambda_0$ twice as large as the one in run T13. Thus, in zero net flux simulations, field amplification depends on box size ($L_{x,z}/\lambda_0$), which is not the case in simulations with net vertical flux.} 
\label{fig:magevocomparenoflux}
\end{center}
\end{figure} 
The MRI saturates at smaller amplitude with no net flux, compared with finite net flux. This can be seen in Figure \ref{fig:magevocomparenoflux}, where the magnetic energy evolution of runs T12 ($v_{A,0}^z/c=1/60$) and T13 ($v_{A,0}^z/c=1/20$) are depicted in black and red lines, respectively. With no net flux, the saturation is no longer characterized by a particular value of $v_A \sim c$ (as in the finite flux case; see Figure \ref{fig:magevocompare}). Instead, different values of $v_{A,0}^z/c$ saturate with similar amplification factors: $B_x/B_{z,0} \approx 10$,  $B_y/B_{z,0} \approx 30$, and $B_z/B_{z,0} \approx 4$. Also, the x-size of the box $L_x/\lambda_0$ appears to play a role in the final saturation, as can be seen by comparing runs T13 and T14 (which are equal except for having $L_x/\lambda_0=8$ and 16, respectively). Indeed, T14 produces somewhat larger values of $B_k^2$ than T13. This can be understood by noting that, for sufficiently large $L_x$, regions of positive and negative initial $B_{z}$ will behave as spatially distinct regions, being able to reach saturation at amplitudes similar to the finite flux cases. Notice also that the earlier saturation suppresses the formation of strong channel flows and magnetic loops (see the density and $B_x$ configurations in the saturated state in Figures \ref{fig:mri2dBxdensnoflux}$e$ and \ref{fig:mri2dBxdensnoflux}$f$).   

\subsection{Pressure Anisotropies and Anisotropic Stresses}
\label{stresses}
The growth of a pressure anisotropy with respect to the local magnetic field can contribute to angular momentum transport via an anisotropic pressure stress, $A_{xy,j}$ $\equiv -\Delta p_j B_xB_y/B^2$, where $\Delta p_j \equiv p_{\perp,j} - p_{||,j}$ and $j$ stands for ions and electrons \citep{QuataertEtAl02,SharmaEtAl06}. This pressure anisotropy is expected in regions where $\vec{B}$ is being amplified by the MRI, due to the adiabatic invariance of the magnetic moment of the particles, $\mu_j \equiv p_{\perp,j}/\rho_j B$. The anisotropic stress may be comparable to the Maxwell stress, $M_{xy} \equiv - B_xB_y/4\pi$ in low-collisionality accretion disks, as was found by previous non-linear studies of the MRI in collisionless plasmas \citep{SharmaEtAl06,SharmaEtAl07}. These studies used a fluid based approach, which did not evolve the pressure in an entirely self-consistent way. Instead, an approximate model was used both to close the fluid equations, and to limit the growth of $\Delta p_j$.  As mentioned in \S \ref{sec:intro}, $\Delta p_j$ is regulated by plasma microinstabilities (mirror, ion cyclotron, electron whistler, etc) acting on scales comparable to the gyroradii of the different species. This provides a mechanism for pressure isotropization in the absence of Coulomb collisions. The effect of these instabilities on particles' velocities is kinetic in nature and can not be consistently captured by a fluid approach. Thus, to date, their effect has been modeled by imposing a ``hard wall" upper limit to $|p_{\perp,j}/p_{||,j} - 1|$, based on the assumption that $\Delta p_j/p_{||,j}$ will grow only until the relevant microinstabilities reach their instability threshold. This criterion is motivated by solar wind observations \citep{BaleEtAl09}, and theoretical and PIC studies of the relevant instabilities \citep[e.g.,][]{GaryEtAl97}. However, how these criteria apply given the simultaneous driving of the MHD turbulence by the MRI remains to be clarified. In this section, we describe the evolution of the anisotropy stress self-consistently using PIC simulations of the MRI, and quantify their contribution to transport in the disk. In \S \ref{sec:2d} we provide a detailed 2D description of pressure anisotropies and their corresponding anisotropic stress $A_{xy,j}$. In \S \ref{sec:1d} we analyze the dependence of $\Delta p_j$ and $A_{xy,j}$ on different simulation parameters using volume-averaged quantities. Finally, in \S \ref{sec:mirror} we illustrate the growth of anisotropy-driven microinstabilities by identifying and analyzing the properties of the relevant small scale modes.

\subsubsection{Spatial distribution of the anisotropies}
\label{sec:2d}
\begin{figure*}
\begin{center}
\centering\includegraphics[width=17cm]{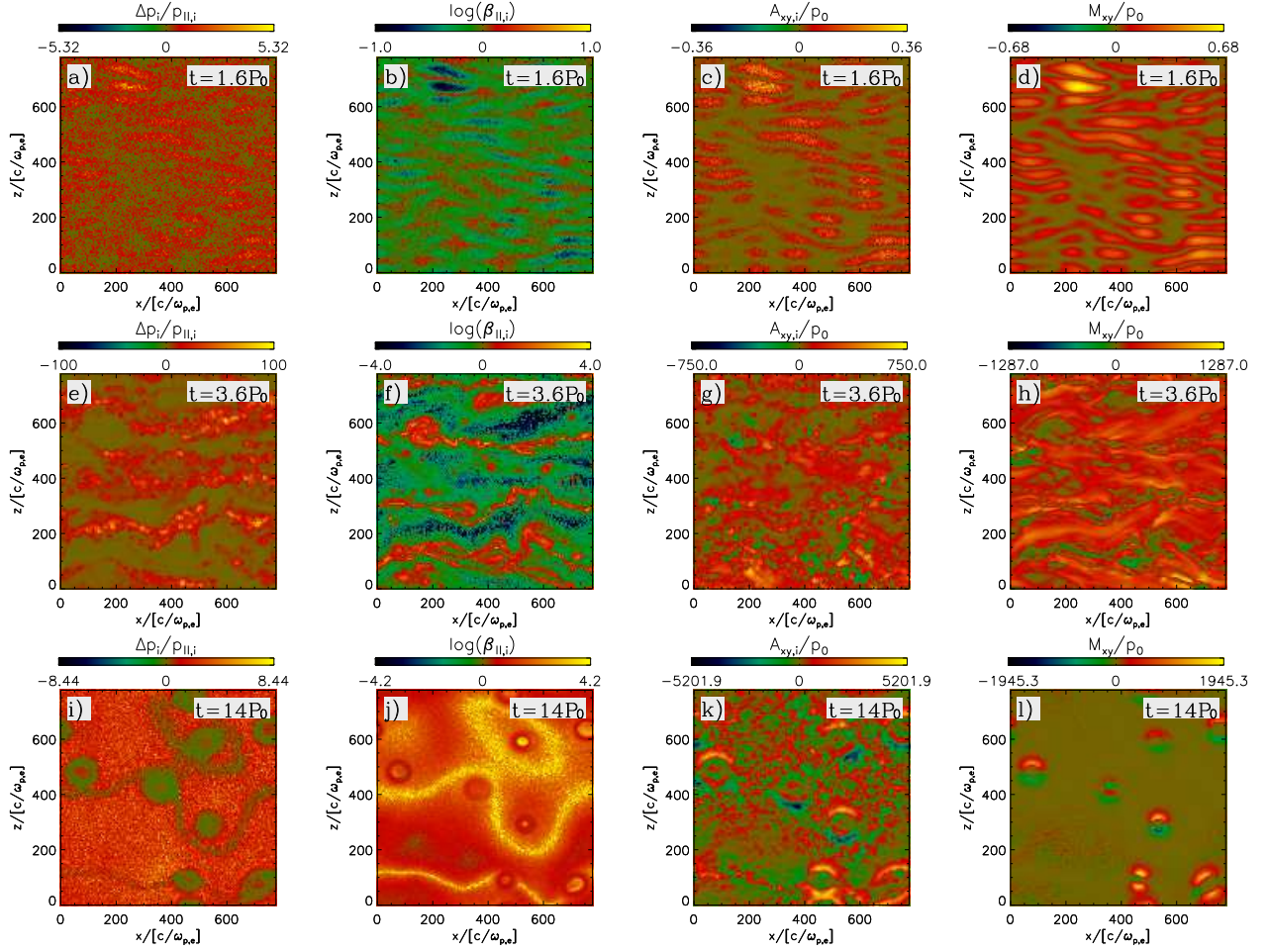}
\caption{2D plots of the ion pressure anisotropy $\Delta p_i/p_{||,i}$ ($\Delta p_i = p_{\perp,i}-p_{||,i}$), the ion parallel beta $\beta_{||,i}$, the ion anisotropic stress $A_{xy,i}/p_0$, and the Maxwell stress $M_{xy}/p_0$, at $t=1.6P_0, 3.6P_0$, and $14P_0$ for run T1 (where $p_0$ is the initial pressure in the plasma). Overall, the pressure anisotropy increases as the MRI grows. At late times, however, the pressure is roughly isotropic in the magnetic loops but anisotropic elsewhere.}
\label{fig:mri2dDelbetstres}
\end{center}
\end{figure*}
\begin{figure}
\begin{center}
\centering\includegraphics[width=8.5cm]{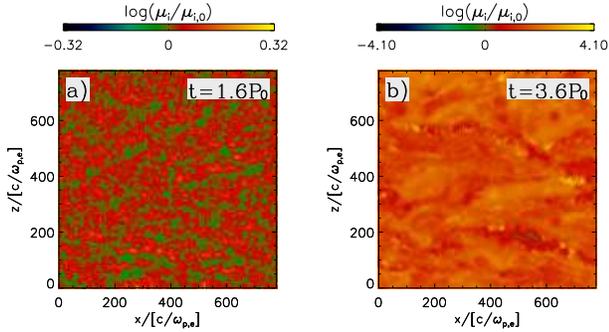}
\caption{2D plots of the average magnetic moment of ions $\mu_i$ ($\equiv p_{\perp,i}/(\rho_iB)$) at $t=1.6P_0$ and $3.6P_0$ for run T1, normalized in terms of its initial value $\mu_{i,0}$. At late times, the magnetic moment has increased significantly, consistent with the isotropization of the plasma pressure by small-scale kinetic instabilities (see Figures \ref{fig:mirror-zoom1} and \ref{fig:mirror-zoom2}).}
\label{fig:mri2dmu}
\end{center}
\end{figure}

 
Figure \ref{fig:mri2dDelbetstres} shows the spatial distribution of the ion pressure anisotropy (Figure \ref{fig:mri2dDelbetstres}$a$) and the corresponding anisotropic stress (Figure \ref{fig:mri2dDelbetstres}$c$) at $t=1.6P_0, 3.6P_0$, and $14P_0$ for run T1. By comparing with Figure \ref{fig:mri2dBxByBzB2}$d$, we see that at $t=1.6 P_0$ the maximum anisotropy occurs in regions of large magnetic amplification, which consequently coincide with minima in $\beta_{||,i}$ (shown in Figure \ref{fig:mri2dDelbetstres}$b$). The anisotropy $\Delta p_i/p_{||,i}$ is also well correlated with the Maxwell stress $M_{xy}$ (Figure \ref{fig:mri2dDelbetstres}$d$) and, therefore, with the anisotropic stress, $A_{xy,i}$. At this early (linear) stage, the ion anisotropy satisfies $\Delta p_i/p_{||,i} \ll 1$, thus plasma microinstabilities are not expected to provide significant pressure isotropization (given that, as we will see below, microinstabilities isotropize the plasma pressure efficiently when $\Delta p_i/p_{||,i} \sim 1/\beta_{||,i}^q$, with $q \sim 1$, and initially $\beta_{i}^z=1$). The lack of ion isotropization can be seen in Figure \ref{fig:mri2dmu}$a$, which shows that at early times the average magnetic moment of ions, $\mu_i$, remains very close to its initial value $\mu_{i,0}$.  \newline

At $t=3.6P_0$, on the other hand, the correlation between $A_{xy,i}$ and $M_{xy}$ gets significantly suppressed. $A_{xy,i}$ is especially suppressed in regions of large $M_{xy}$, which coincide with the regions of the lowest $\beta_{||,i}$. This suggest the presence of an efficient mechanism for pressure anisotropization at low $\beta_{||,j}$. We will quantitatively discuss (in \S \ref{sec:1d}) the most likely mechanism for ion pitch angle scattering in these low $\beta_{||,j}$ regions. Magnetic reconnection is also expected to reduce $\Delta p_j$ at this stage. Indeed, as we will see below, at this time reconnection contributes significantly to particle energization. Since this process is not expected to preserve $\mu_j$, pressure anisotropies must also be reduced due to reconnection. The non-conservation of $\mu_j$ can be seen explicitly in Figure \ref{fig:mri2dmu}$b$, which shows that at late times $\mu_i/\mu_{i,0} \approx 60$ on average (in accordance with $B/B_0 \approx 60$ and $\beta_{i,\perp} \approx 1$).\newline 

At $t=14P_0$ there are no regions in the box where $\beta_{||,i} \ll 1$, which suggests that the correlation between $M_{xy}$ and $A_{xy,i}$ must be to some degree recovered. However, this does not occur. A significant fraction of the magnetic field energy is contained in loops, but the pressure anisotropy within these loops appears to be almost zero. This is consistent with the fact that no significant magnetic amplification occurs in the loops, so they do not develop pressure anisotropies due to $\mu$ conservation. Indeed, loops are a byproduct of the magnetic reconnection of the MRI-amplified field, so no significant growth of $\Delta p_i/p_{||,i}$ is expected to occur in these regions. Thus, we see from Figures \ref{fig:mri2dDelbetstres}$k$ and \ref{fig:mri2dDelbetstres}$l$ that, whereas the largest contribution to $M_{xy}$ comes from the inner part of the loops, the largest magnitudes of $A_{xy,i}$ come from their outer parts. We also note that both $M_{xy}$ and $A_{xy,i}$ can get large negative and positive values, implying that these quantities may produce a negative stress on average. This is because, as can be seen from Figures \ref{fig:mri2dBxByBzB2}$i$ and \ref{fig:mri2dBxByBzB2}$j$, the magnetic loops do not develop an anticorrelation between $B_x$ and $B_y$, as is the case with the MRI channel modes. As we will see below, this produces negative volume-averaged $M_{xy}$ and $A_{xy,j}$ in the late stages of some of our simulations. This is, we believe, an artifact of how our 2D simulations saturate. 

\subsubsection{Volume-averaged anisotropies and stresses}
\label{sec:1d}
\begin{figure}
\begin{center}
\centering\includegraphics[width=9cm]{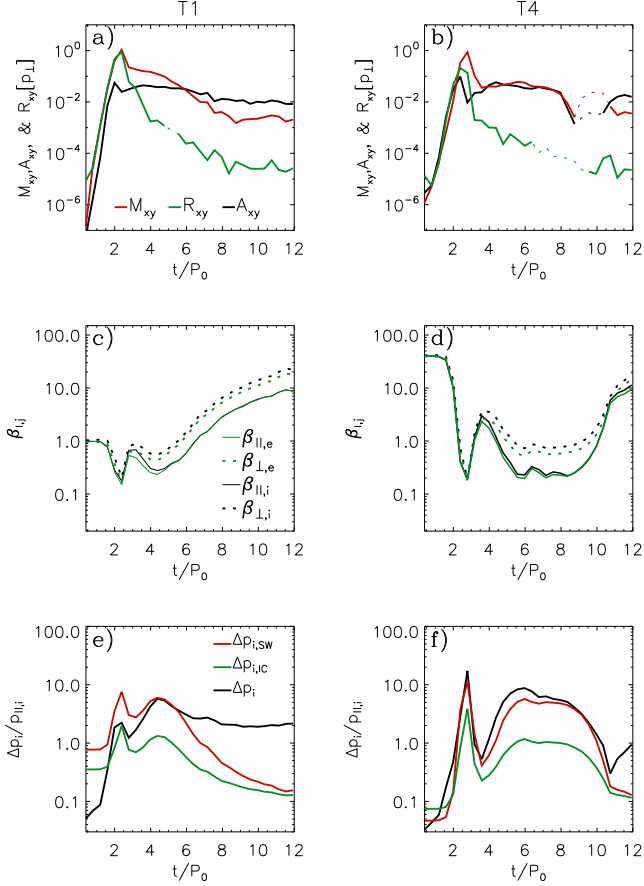}
\caption{The evolution of volume-averaged quantities for simulations T1 and T4. The first row shows the different stresses $M_{xy}$ (red), $R_{xy}$ ($R_{xy} = R_{xy,i} + R_{xy,e}$; green), and $A_{xy}$ ($A_{xy} = A_{xy,i} + A_{xy,e}$; black), normalized in terms of the perpendicular plasma pressure ($p_{\perp} = p_{\perp,i} + p_{\perp,e}$). The second row shows the parallel and perpendicular betas (in solid and dotted lines, respectively), for ions [black] and electrons [green]. These quantities are calculated dividing the volume averages of the particles' pressures and the magnetic field pressure.
The third row shows the pressure anisotropy $\Delta p_i/p_{||,i}$, of ions (black line; electrons follow a qualitatively similar trend), along with the theoretically estimated thresholds for the ion-cyclotron (IC) instability ($\Delta p_{i,IC}$; green) and an empirical threshold obtained from solar wind ion anisotropy measurements ($\Delta p_{i,SW}$), shown by the red line \citep{BaleEtAl09}. Note that the plasma remains near the solar wind threshold (which is very similar to the threshold for excitation of mirror modes), particularly in the higher beta simulation T4 (right column).}
\label{fig:enstressani1}
\end{center}
\end{figure} 
In this section we quantitatively analyze the physics of pressure anisotropy evolution and the corresponding anisotropic stresses, using volume-averaged quantities. Figure \ref{fig:enstressani1} shows the time evolution of volume-averaged stresses, plasma betas, and ion pressure anisotropies for runs T1 and T4. The first column concentrates on run T1, with panel \ref{fig:enstressani1}$a$ depicting the Maxwell stress ($M_{xy}$; red), the Reynolds stress ($R_{xy}=R_{xy_i}+R_{xy,e}$; green), and the anisotropic stress ($A_{xy}=A_{xy,i} + A_{xy,e}$; black). These stresses are plotted as solid (dotted) lines when they are positive (negative), and are normalized in terms of the total perpendicular pressure $p_{\perp}=p_{\perp,i}+p_{\perp,e}$. They are thus an estimate of the contribution to the effective $\alpha$ parameter of \cite{ShakuraEtAl73}. We see that in the exponential growth regime (from $t=0$ to $t\approx2P_0$) $M_{xy}$ and $R_{xy}$ almost coincide, in accordance with the expected linear MRI behavior. On the other hand, $A_{xy}$ appears to grow exponentially at a rate larger than that of $M_{xy}$ and $R_{xy}$. This is because $A_{xy,j}/M_{xy} \sim \Delta p_j/B^2$, so in the linear regime this ratio grows as $\Delta p_j/B_{z,0}^2$. Also, this ratio can be expressed as $A_{xy,j}/M_{xy} = \Delta p_j \beta_{||,j}/(2p_{||,j})$. Thus, since in the linear regime $\Delta p_j/p_{||,j} \ll 1$ and $\beta_{||,j}=1$ (for run T1), the ion and electron anisotropic stresses $A_{xy,j}$ must satisfy $A_{xy,j} \ll M_{xy}$.\newline

After $t \approx 1.5 P_0$, the growth of the anisotropic stress $A_{xy}$ becomes significantly slower than that of the Maxwell stress $M_{xy}$. This can be understood in terms of the dependence of $\Delta p_j/p_{||,j}$ on $\beta_{||,j}$. Figure \ref{fig:enstressani1}$e$ shows the time evolution of the ion pressure anisotropy $\Delta p_i/p_{||,i}$ for run T1 (black line; the electrons follow qualitatively the same trend). When $\beta_{||,i} \lesssim 0.3$ ($t \lesssim 3P_0$),  $\Delta p_i/p_{||,i}$ appears to be constrained by the condition for ion-cyclotron instability growth (green line), as used by \cite{SharmaEtAl06}. This threshold is given by
\begin{equation}
\label{eq:ioncy}
\frac{\Delta p_i}{p_{||,i}} \lesssim \frac{0.35}{\beta_{||,i}^{0.42}},
\end{equation}
which implies that $A_{xy,j}/M_{xy} = \Delta p_j \beta_{||,j}/(2p_{||,j}) \lesssim 0.18 \beta_{||,j}^{0.58} \ll 1$ if $\beta_{||,i} \ll 1$. This explains the lack of correlation between $A_{xy,j}$ and $M_{xy}$ in the low $\beta_{||,i}$ regions at $t=3.6P_0$, as seen in Figures \ref{fig:mri2dDelbetstres}$g$ and \ref{fig:mri2dDelbetstres}$h$.\newline

At larger values of $\beta_{||,i}$ ($t \gtrsim 3P_0$), the pressure anisotropy appears to be limited by a condition less stringent than that of the ion-cyclotron instability. Remarkably, the bound on the ion anisotropy in our simulations is consistent with the maximum anisotropy measured in the solar wind \citep{BaleEtAl09}, shown by the red line. This limit is given by
\begin{equation}
\label{eq:solarwind}
\frac{\Delta p_i}{p_{||,i}} \lesssim \frac{0.77}{(\beta_{||,i} + 0.016)^{0.76}}.
\end{equation}
The bound in equation \ref{eq:solarwind} is very similar to the stability condition for the mirror instability, which is given by:
\begin{equation}
\label{eq:ionmi}
\frac{\Delta p_i}{p_{||,i}} \lesssim \frac{1}{\beta_{\perp,i}}.
\end{equation}
Thus both our simulations and the solar wind data suggest that the mirror instability plays the key role controlling the pressure anisotropy when $\beta_{||,i} \gtrsim 0.3$. The significant role of the mirror instability will be confirmed below by directly identifying the presence of mirror modes embedded in the MRI turbulence. 

Figure \ref{fig:enstressani1} shows that the anisotropic stress begins to dominate for $t \gtrsim 6P_0$.  In addition, at late times the pressure anisotropy $\Delta p_i/p_{||,i}$ is significant larger than the limit provided by Equation \ref{eq:solarwind}.  This can also bee seen in Figures \ref{fig:mri2dDelbetstres}$i$ and \ref{fig:mri2dDelbetstres}$j$, which show the spatial distribution of $\Delta p_i/p_{||,i}$ and $\beta_{||,i}$ of run T1 at $t=14P_0$. Indeed, in most of the volume $\beta_{||,i} \approx 20$, while $\Delta p_i/p_{||,i} \approx 4$ (see, for instance, the region centered at (x,z)=(250,500)$c/\omega_{p,e}$), which is significantly larger than what is expected from Equation \ref{eq:solarwind}. This lack of isotropization is probably due to the large value of the ion Larmor radius $R_{L,i}$ at the end of the simulations, which makes the typical wavelength of the mirror instability close to the MRI wavelength. Indeed, in regions of large anisotropy, $R_{L,i} \approx 100 c/\omega_{p,e}$, while the wavelength of the dominant MRI mode is 800$c/\omega_{p,e}$ (see Figure \ref{fig:mri2dBxByBzB2}$j$). Thus, given the similarity between mirror and MRI wavelengths, we do not expect the mirror modes to grow as effectively as they do in the regime where these scales are well separated.\footnote{Since the ratio of the MRI growth rate and the ion cyclotron frequency in our simulations is $\omega_0/\omega_{c,i} \sim 0.1-0.01$, the mirror and ion-cyclotron time scales are much closer to the MRI time scale than in reality (where $\omega_0/\omega_{c,i} \sim 10^{-7}$). However, our linear calculations show that the pressure anisotropy thresholds for a much larger growth rate are only a factor of $\sim 2$ larger than in Equation \ref{eq:solarwind} and can only partially explain the large departure from the threshold seen in Figure \ref{fig:enstressani1}$e$.}\newline

The large beta behavior of the ion anisotropies can be seen in the second column of Figure \ref{fig:enstressani1}, which shows results for run T4, which initially has $\beta_{j}^z=40$. We see that initially $A_{xy}$ becomes comparable to $M_{xy}$ during the stage of exponential growth. Notice that $A_{xy}$ is similar to $M_{xy}$ only until the end of the exponential growth regime. After that, the significant decrease in $\beta_{||,i}$ makes $M_{xy}$ the dominant stress. We also see in Figure \ref{fig:enstressani1}$b$ that $M_{xy}$ and $A_{xy}$ may acquire negative values. This indicates, as seen in \S \ref{sec:2d}, that in the post-saturation state, stresses may be dominated by loop-like structures where $B_x$ and $B_y$ are not necessarily anti-correlated (as in the case of MRI modes). \newline
Figure \ref{fig:enstressani1}$f$ shows that the maximum of $\Delta p_i/p_{||,i}$ is determined by Equation \ref{eq:solarwind} almost all the time (except at the end of the run, as in run T1). This indicates that using a large initial $\beta_{||}^z$ ensures that isotropization will be dominated by the mirror instability even at the end of the exponential growth, with the ion-cyclotron instability playing a less important role. This large beta behavior also happens for $\beta_{||}^z=10$, as in runs T8 and 11. An interesting question is whether using $\beta_j^z$ significantly larger than 40 would make $A_{xy,i}$ the dominant viscous stress in the MRI-driven turbulence, because the precise value of $A_{xy,i}$ depends on the exact dependence of $\Delta p_i/p_{||,i}$ on $\beta_{||,i}$. Testing this possibility requires using values of $\omega_{c,i}^z/\omega_0$ significantly larger than the ones used here (in order to keep the ion Larmor radii much smaller than the dominant MRI wavelength, $\lambda_0$). This possibility will be investigated in a future work.\newline

We also studied the dependence of $\Delta p_{j}/p_{||,j}$ and the plasma stresses on other simulation parameters. We tested the dependence on: $m_i/m_e$ (with run T2; using $m_i/m_e=5$), different values of  $v_{A,0}^z/c$ (with run T3 and T7; using $v_{A,0}^z/c=1/60$ and 1/120, respectively), initial finite azimuthal flux with $B_{y,0}=B_{z,0}$ (run T5), and a larger magnetization $\omega_{c,i}^z/\omega_0=22$ (run T6). All these simulations (which have $\beta_{||,j}=1$ as in run T1) have the same behavior as run T1, showing that the physics of pressure isotropization obtained in our  simulations is reasonably well converged, with $\beta_{||,j}$ being the only relevant parameter. \newline
Finally, we have also tested numerical convergence using run T9 and T10. Run T9 tested box size dependence by using $L_x=L_z=4\lambda_0$ (half the values used in T1), and run T10 tested space, time, and particle resolution by using $c/\omega_{p,e}/\Delta_x=10$ and $N_{ppc}=6$. No difference with respect to the results of run T1 were found.

\subsubsection{Mirror mode analysis}
\label{sec:mirror}
\begin{figure}
\begin{center}
\centering\includegraphics[width=6.5cm]{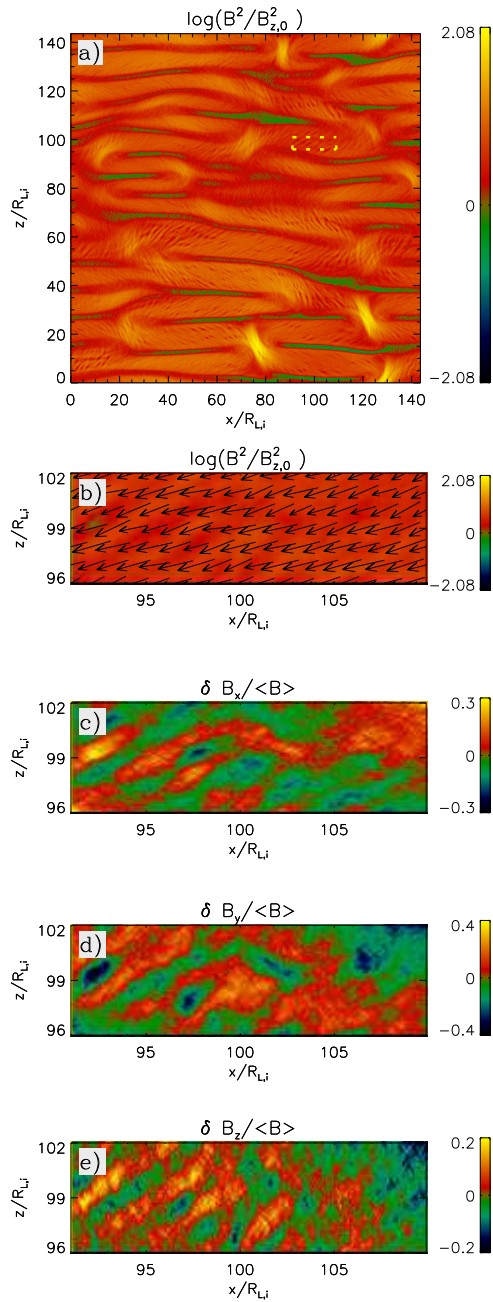}
\caption{Magnetic energy and magnetic field fluctuations for run T8 highlighting the appearance of mirror modes at $t=2P_0$, driven by plasma pressure anisotropies. Panel $a$ shows
$\log(B^2/B_{z,0}^2)$ and panel $b$ shows the same quantity but in a zoomed region centered at $x/R_{L,i} = 100$ and $z/R_{L,i} = 99$ (marked by the yellow-dotted rectangle in panel $a$). Panels $c$, $d$, and $e$ depict $\delta B_k/<B>$, with $k=x, y,$ and $z$. $R_{L,i}$ is the average ion Larmor radius and $< >$ represents volume average within the zoomed-in region. The $x$ and $z$ axes are normalized by $R_{L,i}$ because this is the characteristic wavelength of the fastest growing mirror modes.}
\label{fig:mirror-zoom1}
\end{center}
\end{figure}

\begin{figure}
\begin{center}
\centering\includegraphics[width=6.5cm]{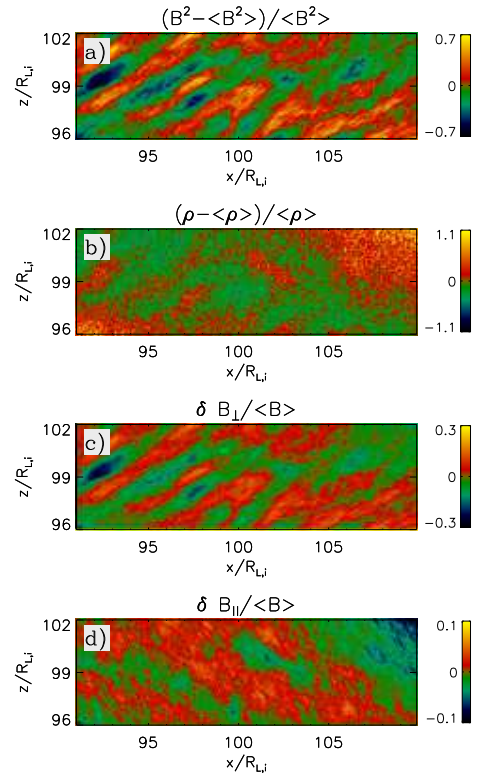}
\caption{The fluctuations in magnetic energy and plasma density for the same region depicted in Figure \ref{fig:mirror-zoom1} are shown in panels $a$ and $b$, respectively. The anti-correlation between these two quantities is a signature of the mirror instability. Panels $c$ and $d$ show magnetic field fluctuations that are perpendicular and parallel to the dominant mirror wavevector $\vec{k}$, respectively. The components parallel to $\vec{k}$ have an amplitude smaller than the perpendicular ones, showing that these modes roughly satisfy the mirror polarization $\delta \vec{B} \perp \vec{k}$.}
\label{fig:mirror-zoom2}
\end{center}
\end{figure}

As seen in \S \ref{sec:1d}, the physics of pressure isotropization at $\beta_{||,j} \gtrsim 0.3$ appears to be well described by the mirror instability, both for ions and (the large mass) electrons. In this section, we analyze the structure of the small scale modes in the plasma, and check whether they satisfy mirror mode properties. Figure \ref{fig:mirror-zoom1} shows the case of run T8, which is analogous to T1 but with $\beta_{j}^z=10$ (the larger $\beta_j^z$ makes the effects of the mirror mode more prominent, as can be seen in Figure \ref{fig:enstressani1}$f$). Plot \ref{fig:mirror-zoom1}$a$ shows $\log(B^2/B_{z,0})$ at $t=2P_0$. It is clear visually that, in regions of amplified magnetic field, small scale fluctuations arise. The length scale of these modes is about $2 R_{L,i}$ (where $R_{L,i}$ is the average Larmor radius of the ions), as can be seen from plots \ref{fig:mirror-zoom1}$b$- \ref{fig:mirror-zoom1}$e$. These plots depict  $\log(B^2/B_{z,0})$ and $\delta B_j/<B>$ in a zoomed region marked by the small rectangle centered at $x/R_{L,i}=100$ and $z/R_{L,i}=99$ in plot \ref{fig:mirror-zoom1}$a$ (where $\delta B_k = B_k - <B_k>$ is a measure of the magnetic field fluctuations along the $k-$axis, and $< >$ represents volume average within the zoomed-in region). The typical wavelength of $\sim 2 R_{L,i}$ of the mirror modes is also observed in simulation T4, where the initial ion temperature is four times that of simulation T8. Also, we confirmed the numerical convergence of this scaling using run T11, which has a spatial resolution of $c/\omega_{p,e}=14 \Delta_x$ (twice the one of run T8).\newline
One of the properties of the mirror instability is the anti-correlation between $B^2$ and plasma density $\rho$. This anti-correlation is present in Figures \ref{fig:mirror-zoom2}$a$ and \ref{fig:mirror-zoom2}$b$, which depict $(B^2-<B^2>)/<B^2>$ and $(\rho-<\rho>)/<\rho>$ in the same zoomed-in region shown in Figure \ref{fig:mirror-zoom1}. Also, mirror modes satisfy $\delta \vec{B} \perp \vec{k}$. This can be checked by computing the components of $\delta \vec{B}$ parallel and perpendicular to $\vec{k}$, shown in plots \ref{fig:mirror-zoom2}$c$ and \ref{fig:mirror-zoom2}$d$, respectively (where $\vec{k}$ and $\hat{x}$ are estimated to form an angle of $120^\circ$). The amplitude of $\delta B_{\perp}$ is significantly larger than that of $\delta B_{||}$, consistent with the mirror mode polarization. Finally, we can compare the projection of $\delta \vec{B}$ on the $x-z$ plane (shown by the $\delta B_x$ and $\delta B_z$ plots in Figures \ref{fig:mirror-zoom1}$c$ and \ref{fig:mirror-zoom1}$e$) with the analogous projection of $\vec{B}$ (shown by black arrows in Figure \ref{fig:mirror-zoom1}$b$). The magnetic fluctuations in the $x-z$ plane roughly satisfy $\delta \vec{B} || \vec{B}$, another polarization property of the mirror modes. 

\subsection{The Energy Distribution of Particles}
\label{heating}
\begin{figure}
\begin{center}
\centering\includegraphics[width=8cm]{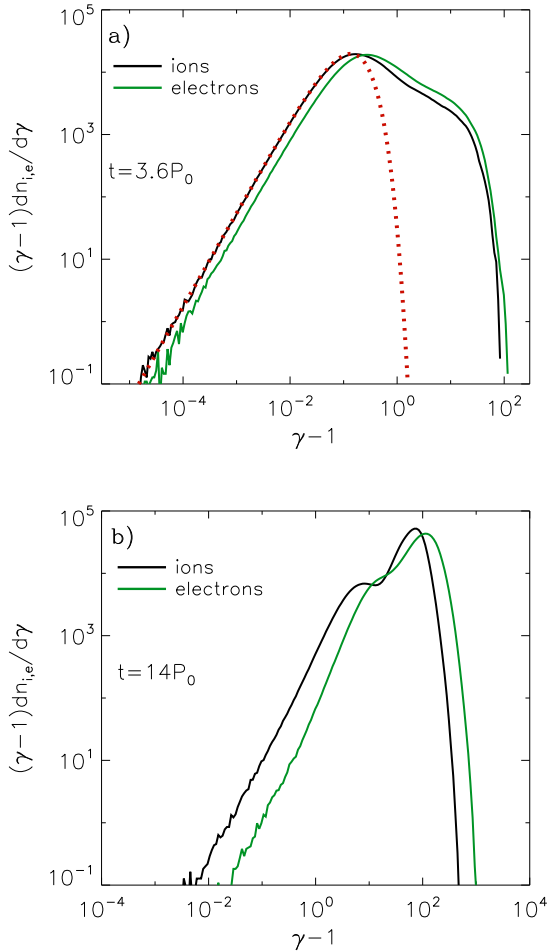}
\caption{The particle spectra for run T1 at times $t=3.6P_0$ (panel $a$) and $t=14P_0$ (panel $b$). Black and green lines correspond to ions and electrons, respectively. While the early-time spectra are composed of a thermal (shown for reference in red-dotted line) and power-law distribution, the late-time spectra correspond to a two-temperature distribution for each species. The formation of the power-law distribution is clearly correlated with magnetic reconnection, while the heating at later times appears to be dominated by viscous heating due to the anisotropic stress.}
\label{fig:spectra}
\end{center}
\end{figure}

In this section we explore the signatures of the different heating processes in the energy spectra of the particles by analyzing the case of run T1. Figures \ref{fig:mri2dBxByBzB2} and \ref{fig:mri2dBxdens} show that at $t=3.6P_0$ efficient migration into longer wavelength MRI modes is happening, with a correspondingly significant rate of magnetic field energy dissipation through reconnection. Figure \ref{fig:spectra}$a$ shows that the average ion and electron spectra at that moment are composed of a thermal distribution, plus a power-law tail with spectral index of $\sim 1.5$ \footnote{Notice that this is the same reconnection-driven spectral index found by \cite{SironiEtAl11} in their PIC simulations of relativistic striped wind shocks. This suggests a possible connection between relativistic and non-relativistic reconnection-driven non-thermal particle spectra.} (black and green correspond to ions and electrons, respectively).  On the other hand, Figure \ref{fig:spectra}$b$ shows the energy spectra at $t=14P_0$, which corresponds to the ``quiescent" state, where no reconnection happens whatsoever. We see that, at this stage, there is not a power-law tail and, instead, a high-energy bump appears. The high-enegy particles concentrate in the regions outside of the magnetic loops, where $A_{xy,j}$ is relatively large, as can be seen from Figure \ref{fig:mri2dDelbetstres}$k$. This indicates 
that at later times particle energization occurs mainly via a viscous heating proportional to $A_{xy,j}$ \citep[like the one suggested in equation 6 of ][]{SharmaEtAl07}. The energy spectra between $t=3.6P_0$ and $t=14P_0$ are a combination of these two distributions. In a future study, we will explore significantly larger values of $m_i/m_e$ to shed light on the different heating mechanisms for ions and electrons in collisionless disks.   

\section{Summary and Discussion}
\label{sec:disconclu}
In this paper we have studied the magnetorotational instability (MRI) in a collisionless plasma using first-principles particle-in-cell (PIC) simulations. Our motivation is the application to low accretion rate, radiatively inefficient accretion flows (e.g., \citealt{Narayan1998}). These flows are expected to be present in systems accreting at less than a few percent of the Eddington rate.  This includes the central black hole of our Galaxy (Sgr A*) and most nearby galaxies, as well as the low-hard state of X-ray binaries. \newline

In the first part of the paper (\S \ref{sec:injection}), we studied the linear dynamics of the MRI using 1D simulations in which only wavevectors along the z-direction -- the rotation axis -- are resolved.  We focused in particular on understanding the effects of the plasma magnetization $\omega_{c,i}^z/\omega_0$ (the ratio of the initial ion cyclotron frequency to the disk rotation frequency) and finite particle temperature on the dispersion relation of the MRI. Our results were in reasonable agreement with previous analytical studies of the collisionless MRI \citep{QuataertEtAl02, KrolikEtAl06, Ferraro07}. \newline

In the second part of the paper, we studied the multidimensional aspects of the MRI using local 2D (axisymmetric) simulations.  Given the computational effort involved in PIC simulations, we defer fully 3D simulations to future work.  In our 2D calculations, we focused on the evolution of the plasma pressure, magnetic energy, and plasma stresses, and assessed their dependence on the different physical and simulation parameters.  We found that the overall evolution of the MRI in axisymmetric simulations is qualitatively similar to that found in previous MHD simulations and simulations that included fluid models of kinetic effects \citep{SharmaEtAl06}.  In particular, in the PIC simulations with net magnetic flux, the MRI only saturates when the Alfv\'en speed becomes comparable to the speed of light (irrespective of the initial value of the Alfv\'en speed).  This is consistent with the absence of saturation of MRI channel models in analogous axisymmetric fluid simulations (see Figure \ref{fig:t1mhd2d3d}). By contrast, for simulations with no net magnetic flux, the MRI saturates at a lower amplitude before $v_A \sim c$ (see Figure \ref{fig:magevocomparenoflux}).\newline

By adiabatic invariance of the magnetic moment of particles, the amplification of the magnetic field by the MRI in a collisionless plasma produces pressure anisotropies, $\Delta p_j=p_{\perp,j}-p_{||,j}$, for each species $j$ (where the directions are measured relative to the local magnetic field).  These anisotropies can be important for angular momentum transport and particle heating, since they give rise to an anisotropic pressure stress, $A_{xy,j} \equiv -B_xB_y \Delta p_j/B^2$, that may be comparable to the magnetic stress \citep{QuataertEtAl02, SharmaEtAl06}.  This is effectively a macroscopic viscosity in a collisionless plasma.  We investigated this physics in a self-consistent way by studying the evolution of the plasma pressure during the linear and non-linear phases of the MRI.  We found that the importance of the anisotropic stress depends on the instantaneous plasma $\beta$. For small $\beta_i \lesssim 10$, the ion anisotropic stress, $A_{xy,i}$, is smaller than the Maxwell stress $M_{xy}$. However, $A_{xy,i}$ may be comparable or even surpass $M_{xy}$ when $\beta_{i} \gtrsim 10$. This regime is difficult to achieve in our simulations due to the continuous build-up of $\vec{B}$, which grows until $v_{A}^x$ and $v_A^z \sim c$, and $v_A^y \sim 10c$. This difficulty could in principle be overcome by using large initial $\beta_j$. However, this would require using values of the magnetization parameter $\omega_{c,i}^z/\omega_0$ significantly larger than those used here (so that the MRI and microinstabilities length scales are properly separated), which is computationally difficult (see Equation \ref{eq:scale}). 
We expect that the saturation of both the magnetic field and the pressure anisotropy will be fundamentally different in 3D simulations. In that case, magnetic field amplification is expected to stop when $v_A \ll c$, which would allow the existence of a turbulent state in which $\beta_{||,i} \gtrsim 10$ (see, e.g., the 2D vs. 3D fluid simulations in Figure \ref{fig:t1mhd2d3d}).  Our results suggest that transport and heating due to the anisotropic stress will be important in this regime.  \newline

Absent any mechanism of temperature isotropization-- which is provided by collisions in the fluid limit -- the pressure anisotropy would continue to grow throughout the linear phase of the MRI in a low-collisionality plasma.  In our calculations, we find direct evidence for collisionless processes that provide temperature isotropization in the absence of Coulomb collisions.  In particular, the $p_\perp > p_\parallel$ state that is generically created by the MRI is unstable to the excitation of mirror modes (see Figures \ref{fig:mirror-zoom1} and \ref{fig:mirror-zoom2}).  Moreover, once mirror excitation commences the volume-averaged pressure anisotropy remains near the threshold for the onset of the mirror instability, particularly in our higher $\beta_j^z$ simulations (as shown in Figures \ref{fig:enstressani1}$e$ and Figures \ref{fig:enstressani1}$f$).  This is consistent with the evolution of the pressure anisotropy observed in the near-Earth solar wind \citep{BaleEtAl09}.  Our results, analogous theoretical work in the solar wind context \citep{HellingerEtAl06}, and the solar wind measurements all strongly suggest that temperature isotropization in moderate $\beta$ low-collisionality plasmas is dominated by (reasonably) well understood velocity space instabilities (in particular, the mirror, firehose, and ion cyclotron instabilities).\newline

In future work, we will study the heating of particles in MRI turbulence in detail, paying particular attention to how the assumed electron to ion mass ratio $m_e/m_i$ gives rise to different electron and ion heating physics. In our preliminary analysis in the present paper, we found that particle heating occurs via two dominant mechanisms: one is reconnection and the other is the viscous heating produced by the anisotropic stress $A_{xy,j}$ \citep{SharmaEtAl07}. These different heating processes are expected to imprint different signatures in the energy spectra of particles.  Perhaps most interestingly, we find that magnetic reconnection produces a distinctive power-law distribution function (with spectral index $\approx 1.5$) as the MRI becomes nonlinear, but before the turbulence has saturated and died away (see Figure \ref{fig:spectra}$a$). This spectral index is similar to the one found by \cite{SironiEtAl11} in their PIC study of reconnection forced by relativistic shocks in the striped wind of pulsars. Whether this non-thermal energization is due to the reconnection electric field parallel to the current sheet \citep[as in the case of][]{SironiEtAl11}, or by a Fermi-like process due to the formation of magnetic loops \citep{DrakeEtAl06}, will be clarified in detail in a future work. The importance of reconnection diminishes once the MRI saturates.   At that point, the particle energization appears to be dominated by the anisotropic stress $A_{xy,j}$, which is largest outside of the magnetic loops. This process produces a high-energy bump in the energy spectra of particles (see Figure \ref{fig:spectra}$b$), with the most energetic particles concentrating outside of the magnetic loops. In our calculations we do not find unambiguous evidence for heating via the turbulent cascade of Alfv\'en and slow waves that is expected to be set up via the nonlinear saturation of the MRI \citep{QG1999}.  This could be a limitation of our 2D simulations which do not develop sustained turbulence; in addition, such a cascade may be difficult to resolve given the modest dynamic range in our simulations between the inner and outer turbulent scales.
\newline

The PIC simulations presented in this paper have numerous advantages relative to fluid calculations for studying the physics of the MRI in low-collisionality plasmas.  In particular, our PIC simulations represent the first self-consistent study of magnetic field amplification and saturation, particle heating, and pressure anisotropy evolution in MRI-driven turbulence.  There are, however, also drawbacks associated with PIC simulations.  In addition to being very computationally demanding, the need to resolve the electron skin depth implies that there is always an unphysically small dynamic range between the initial ion cyclotron frequency $\omega_{c,i}$ (where $i$ stands for `ions') and the disk rotation frequency $\omega_0$: our simulations initially have $\omega_{c,i}/\omega_0\sim 10-100$ instead of $\omega_{c,i}/\omega_0 \sim 10^7$ in real systems. This ratio, however, increases to $\omega_{c,i}/\omega_0 \sim100-1000$ in the nonlinear regime, considering both the magnetic field amplification and the growth of the mean Lorentz factor of the particles. We find no strong dependence of our results on the initial value of $\omega_{c,i}/\omega_0$, suggesting that we are adequately in the ``MHD'' regime, but this will need to be carefully studied in 3D simulations as well, where the computational requirements are even more demanding.\newline

Another limitation of the current calculations is that our results are strictly valid only in the limit $v_0 \ll c$ (where $v_0$ is the bulk rotation velocity of the plasma).  In particular, the PIC equations evolved here neglect additional terms in Maxwell's equations that arise from being in a rotating reference frame (see Appendix \ref{sec:shearcoord}); these are self-consistently small in the limit of small rotation velocities.  Our analysis does consistently allow for relativistic temperatures and thus it is possible to study quantitatively the limit of relativistic electrons, but non-relativistic ions that is of particular interest for radiatively inefficient accretion flow models.



\acknowledgements 
This work was supported in part by NSF-DOE Grant PHY- 0812811.  Support for P.S. was provided by NASA through the Chan- dra Postdoctoral Fellowship grant PF8-90054 awarded by the Chandra X-Ray Center, which is operated by the Smithsonian Astrophysical Observatory for NASA under contract NAS8- 03060.  The computations for this paper were perfomed on the Henyey cluster at UC Berkeley, supported by NSF grant AST-0905801. The authors would also like to thank Greg Hammett and Tobi Heinemann for useful discussions.


\appendix

\section{Shearing Coordinates}
\label{sec:shearcoord}
We describe the plasma in the shearing frame, $S'$, which is related to the usual cartesian frame $S$ by the following coordinate transformation:
 \begin{equation}
\begin{array}{rclcccrcl}
x'&=&x,&\textrm{  }&\textrm{  }&\textrm{  }&y'&=&\Gamma(y-vt),\\
z'&=&z,&\textrm{  }&\textrm{ and }&\textrm{  }&t'&=&\Gamma(t-vy/c^2),
\end{array}
\label{eq:lorentz}
\end{equation}
where the primed coordinates correspond to the frame $S'$, $\Gamma= (1- v^2/c^2)^{-1/2}$ is the Lorentz factor due to shear, $v=-sx$ is the $y$-component of the shear velocity, and $s$ is the shear parameter ($s=3\omega_0/2$ in the case of a Keplerian disk). In $S'$ the plasma is initially at rest (with no shearing velocity), which allows us to replace shearing box periodic boundary condition by standard periodic boundary conditions. \newline
The transformation of the electric and magnetic fields from $S$ to $S'$ can be expressed as the standard Lorentz invariant transformation
\begin{equation}
\begin{array}{rclcccrcl}
E'_{y} &=& E_{y}, &&\textrm{  }&&\vec{E}'_{\perp} &=& \Gamma (\vec{E}_{\perp} + \frac{\vec{v}}{c}\times\vec{B}_{\perp}),\\
B'_{y} &=& B_{y}, &&\textrm{and}&& \vec{B}'_{\perp} &=& \Gamma (\vec{B}_{\perp} - \frac{\vec{v}}{c}\times\vec{E}_{\perp}).\\
\end{array}
\label{eq:fields}
\end{equation}
As we will see below, the evolution of the fields in $S'$ is given by modified versions of Maxwell's equations. The modifications arise from the dependence of $v$ and $\Gamma$ on $x$. Although we are interested in the small box limit, we derive the evolution of $\vec{E}'$ and $\vec{B}'$ assuming an arbitrarily large box, with the small box limit taken only at the end of the calculations. Since the module of $v(x)$ can not exceed $c$, we will use a shear profile where this is satisfied regardless of the value of $x$. To do that, we will impose that the ``local" shear profile seen by an observer moving with the flow is $v_{local}=-sx_{local}$, where $x_{local}=0$ at the observer's position. One can show that this condition implies a global shear in the box given that $v/c=-\textrm{arctanh}(sx/c)$. With this space dependence of $v(x)$, the $x-$derivatives of $v$ and $\Gamma$ will be given by
\begin{equation}
dv/dx=-s/\Gamma^2 \textrm{ }\textrm{ }\textrm{ and }\textrm{ }\textrm{ } d\Gamma/dx=-s\Gamma v/c^2,
\label{eq:derivs}
\end{equation}
which we will use in the derivation of the field dynamics described below.

\subsection{{\bf Modifications to Maxwell's equations}}
We determine the changes to each component of Maxwell's equations one by one. Let us start with Faraday's equation.
\subsubsection{{\bf Faraday's equation: x component}}
Given the transformations defined in Equations \ref{eq:lorentz} and \ref{eq:fields}, we want to know how the equation
\begin{equation}
\frac{\partial B_x(\vec{r},t)}{\partial t} = -(c\nabla \times \vec{E}(\vec{r},t))_x 
\end{equation}
is modified in the coordinate system $S'$. From Equations \ref{eq:fields} we get 
\begin{equation}
\begin{array}{rclcccrclcccrcl}
B'_x(\vec{r}',t') &=& \Gamma (B_x(\vec{r},t) - \frac{v}{c}E_z(\vec{r},t)),&&&&E'_z(\vec{r}',t') &=& \Gamma (E_z(\vec{r},t) - \frac{v}{c}B_x(\vec{r},t)),&&\textrm{and}&&E'_y(\vec{r}',t') &=& E_y(\vec{r},t)
\end{array}
\end{equation}
Then, from Equation \ref{eq:lorentz} it is possible to show that 
\begin{eqnarray}
\frac{\partial B_x'(\vec{r}',t')}{\partial t'} &=& \Gamma^2 \Big(\frac{\partial B_x(\vec{r},t)}{\partial t} -\frac{v}{c}\frac{\partial E_z(\vec{r},t)}{\partial t} +v\frac{\partial B_x(\vec{r},t)}{\partial y} - \frac{v^2}{c}\frac{\partial E_z(\vec{r},t)}{\partial y}\Big),\nonumber \\
\frac{\partial E_y'(\vec{r}',t')}{\partial z'} &=& \frac{\partial E_y(\vec{r},t)}{\partial z}, \textrm{ and}\nonumber \\ 
\frac{\partial E_z'(\vec{r}',t')}{\partial y'} &=& \Gamma^2\Big(\frac{v}{c^2}\frac{\partial E_z(\vec{r},t)}{\partial t} -\frac{v^2}{c^3}\frac{\partial B_x(\vec{r},t)}{\partial t} + \frac{\partial E_z(\vec{r},t)}{\partial y} - \frac{v}{c}\frac{\partial B_x(\vec{r},t)}{\partial y}\Big).
\label{eq:inducx}
\end{eqnarray}
Combining these three expressions, one can show that
\begin{equation}
\frac{\partial B_x'(\vec{r}',t')}{\partial t'} = -(c\nabla'\times\vec{E}'(\vec{r}',t'))_x.
\end{equation} 
So the $x-$component of Faraday's equation does not transform from $S$ to $S'$.

\subsubsection{{\bf Faraday's equation: $y$ component}}
Here the procedure is analogous to the case described above. However, since the $y$ component includes partial derivatives with respect to $x'$, the $x'$ dependence of $v$ and $\Gamma$ will give rise to new terms. From Equations (\ref{eq:fields}) we get:
\begin{equation}
\begin{array}{rclcccrclcccrcl}
B'_y(\vec{r}',t') &=& B_y(\vec{r},t),&&&&E'_z(\vec{r}',t') &=& \Gamma (E_z(\vec{r},t) - \frac{v}{c}B_x(\vec{r},t)),&&\textrm{and}&&E'_x(\vec{r}',t') &=& \Gamma (E_x(\vec{r},t) + \frac{v}{c}B_z(\vec{r},t)).
\end{array}
\end{equation}
Then
\begin{eqnarray}
\frac{\partial B_y'(\vec{r}',t')}{\partial t'} &=& \Gamma \Big(\frac{\partial B_y(\vec{r},t)}{\partial t} + v\frac{\partial B_y(\vec{r},t)}{\partial y}\Big), \nonumber \\
\frac{\partial E_x'(\vec{r}',t')}{\partial z'} &=& \Gamma \Big(\frac{\partial E_x(\vec{r},t)}{\partial z} + \frac{v}{c}\frac{\partial B_z(\vec{r},t)}{\partial z}\Big), \textrm{ and} \nonumber \\ 
\frac{\partial E_z'(\vec{r}',t')}{\partial x'} &=& \frac{\partial \Gamma}{\partial x}\Big(E_z(\vec{r},t)-\frac{v}{c}B_x(\vec{r},t)\Big) - \frac{\Gamma}{c} \frac{\partial v}{\partial x}B_x(\vec{r},t) + \Gamma\Big(\frac{\partial E_z(\vec{r},t)}{\partial x} - \frac{v}{c}\frac{\partial B_x(\vec{r},t)}{\partial x}\Big) \nonumber \\&&
+\Gamma \Big(\frac{\partial \Gamma}{\partial x}(y'+vt') + \Gamma t' \frac{\partial v}{\partial x}\Big)\Big(\frac{\partial E_z}{\partial y} - \frac{v}{c}\frac{\partial B_x}{\partial y}\Big)
+\Gamma \Big(\frac{\partial \Gamma}{\partial x}(t'+\frac{vy'}{c^2}) + \Gamma \frac{y'}{c^2} \frac{\partial v}{\partial x}\Big)\Big(\frac{\partial E_z}{\partial t} - \frac{v}{c}\frac{\partial B_x}{\partial t}\Big). 
\label{eq:inducy}
\end{eqnarray}
Combining Equations \ref{eq:inducy}, and using the expressions for the $x$ derivatives of $v$ and $\Gamma$ shown in Equation \ref{eq:derivs}, it is straightforward to obtain
\begin{equation}
\frac{\partial B_y'(\vec{r}',t')}{\partial t'} = -(c\nabla'\times\vec{E}'(\vec{r}',t'))_y - sB_x'(\vec{r}',t') + s\Big(ct'\frac{\partial E_z'}{\partial y'} + \frac{y'}{c}\frac{\partial E_z'}{\partial t'}\Big).
\end{equation} 

\subsubsection{{\bf Faraday's equation: $z$ component}}
The derivation in this case is analogous to the case of the $y$ component. From Equations (\ref{eq:fields}) we can get:
\begin{equation}
\begin{array}{rclcccrclcccrcl}
B'_z(\vec{r}',t') &=& \Gamma (B_z(\vec{r},t) + \frac{v}{c}E_x(\vec{r},t)),&&&&E'_x(\vec{r}',t') &=& \Gamma (E_x(\vec{r},t) + \frac{v}{c}B_z(\vec{r},t)),&&\textrm{and}&&E'_y(\vec{r}',t') &=& E_y(\vec{r},t),
\end{array}
\end{equation}
which imply that
\begin{eqnarray}
\frac{\partial B_z'(\vec{r}',t')}{\partial t'} &=& \Gamma^2 \Big(\frac{\partial B_z(\vec{r},t)}{\partial t} +\frac{v}{c}\frac{\partial E_x(\vec{r},t)}{\partial t} + v\frac{\partial B_z(\vec{r},t)}{\partial y} + \frac{v^2}{c}\frac{\partial E_x(\vec{r},t)}{\partial y}\Big),\nonumber \\
\frac{\partial E_x'(\vec{r}',t')}{\partial y'} &=& \Gamma^2 \Big(\frac{v}{c^2}\frac{\partial E_x(\vec{r},t)}{\partial t} + \frac{v^2}{c^3}\frac{\partial B_z(\vec{r},t)}{\partial t} + \frac{\partial E_x(\vec{r},t)}{\partial y} + \frac{v}{c}\frac{\partial B_z(\vec{r},t)}{\partial y}\Big), \textrm{ and}\nonumber \\ 
\frac{\partial E_y'(\vec{r}',t')}{\partial x'} &=& \frac{\partial E_y(\vec{r},t)}{\partial x} + \frac{\partial E_y(\vec{r},t)}{\partial y}\Big( \frac{\partial \Gamma}{\partial x'}(y'+vt') + \frac{\partial v}{\partial x'}\Gamma t' \Big) + \frac{\partial E_y(\vec{r},t)}{\partial t}\Big( \frac{\partial \Gamma}{\partial x'}(t'+\frac{v}{c^2}y') + \frac{\partial v}{\partial x'}\Gamma \frac{y'}{c^2} \Big).
\label{eq:inducz}
\end{eqnarray}
Then, combining Equations (\ref{eq:inducz}) with the derivatives of $v$ and $\Gamma$ given in \ref{eq:derivs}, it can be shown that 
\begin{equation}
\frac{\partial B_z'(\vec{r}',t')}{\partial t'} = -(c\nabla'\times\vec{E}'(\vec{r}',t'))_z -s \Big(ct'\frac{\partial E_y'(\vec{r}',t')}{\partial y'} + \frac{y'}{c}\frac{\partial E_y'(\vec{r}',t')}{\partial t'}\Big).
\end{equation} 
The three components of Faraday's equation (Equations \ref{eq:inducx}, \ref{eq:inducy}, and \ref{eq:inducz}) can be combined in a single expression:
\begin{equation}
\frac{\partial \vec{B}'(\vec{r}',t')}{\partial t'} = -c\nabla'\times\vec{E}'(\vec{r}',t') -sB_x(\vec{r}',t')\hat{y}  + s \Big(ct'\frac{\partial \vec{E}'(\vec{r}',t')}{\partial y'} + \frac{y'}{c}\frac{\partial \vec{E}'(\vec{r}',t')}{\partial t'}\Big)\times \hat{x}.
\label{eq:faraday}
\end{equation} 
Equation \ref{eq:faraday} contains a time dependent term arising from the time evolution of the $S'$ coordinates with respect to the ones in $S$. Indeed, as time goes on, $\partial \vec{E}'/\partial x'$ ($\partial \vec{B}'/\partial x'$) must increase its difference relative to $\partial \vec{E}/\partial x$ ($\partial \vec{B}/\partial x$) simply because two points of equal $y$ in $S$ ($\Delta y =0$) have a difference in $y'$ ($\Delta y'$) that grows linearly with time. This explains why the time dependent term must be proportional to the extent to which the fields depend on $y'$. Thus, in the 2D case (relevant to this work), this time dependent term does not play any role.\newline 

In addition to the time dependent term, there is the term proportional to $y'$, which appears to be inconsistent with a 2D treatment of equation \ref{eq:faraday}. However, in the small box limit (i.e., when $y'$ is much smaller than the distance from the box origin to the disk center, $r_0$), the magnitude of the velocity $sy'$ is much smaller than the orbital velocity of the disk at $r_0$ ($v_0$). Thus, since $v_0 \ll c$,  $sy'$ must also be $\ll c$, which allows us to neglect the space dependent term in Equation \ref{eq:faraday}, especially if one expects $|\vec{E}'| \ll |\vec{B}'|$ in non-relativistic turbulence. Thus, given our non-relativistic, small box approximation, we can neglect the $y'$ dependence in Equation \ref{eq:faraday}, which allows us to formally use a 2D approach to this problem.\footnote{Using a Galilean instead of Lorentzian transformation of coordinates and fields, it easy to show that the $y'$ dependence in equation \ref{eq:faraday} does not appear, which confirms the validity of our result in the non-relativistic limit.} In that limit, Faraday's equation can be expressed as: 
\begin{equation}
\frac{\partial \vec{B}'(\vec{r}',t')}{\partial t'} = -c\nabla'\times\vec{E}'(\vec{r}',t') - sB_x'(\vec{r}',t')\hat{y}.
\label{ref:inductiontotal}
\end{equation}

\subsection{{\bf Ampere's equation: x component}}
We want to get an equation analogous to
\begin{equation}
\frac{\partial E_x(\vec{r},t)}{\partial t} = (c\nabla \times \vec{B}(\vec{r},t))_x - 4\pi J_x 
\end{equation}
in the coordinate system $S'$. From Equations \ref{eq:fields} we get 
\begin{equation}
\begin{array}{rclcccrclcccrcl}
E'_x(\vec{r}',t') &=& \Gamma (E_x(\vec{r},t) + \frac{v}{c}B_z(\vec{r},t)),&&&&B'_z(\vec{r}',t') &=& \Gamma (B_z(\vec{r},t) + \frac{v}{c}E_x(\vec{r},t)),&&\textrm{and}&&B'_y(\vec{r}',t') &=& B_y(\vec{r},t).
\end{array}
\end{equation}
Thus, from Equations (\ref{eq:lorentz}) we get that 
\begin{eqnarray}
\frac{\partial E_x'(\vec{r}',t')}{\partial t'} &=& \Gamma^2 (\frac{\partial E_x(\vec{r},t)}{\partial t} + \frac{v}{c}\frac{\partial B_z(\vec{r},t)}{\partial t} + v\frac{\partial E_x(\vec{r},t)}{\partial y} + \frac{v^2}{c}\frac{\partial B_z(\vec{r},t)}{\partial y}),\nonumber \\
\frac{\partial B_y'(\vec{r}',t')}{\partial z'} &=& \frac{\partial B_y(\vec{r},t)}{\partial z}, \textrm{ and}\nonumber \\ 
\frac{\partial B_z'(\vec{r}',t')}{\partial y'} &=& \Gamma^2 (\frac{v}{c^2}\frac{\partial B_z(\vec{r},t)}{\partial t} + \frac{v^2}{c^3}\frac{\partial E_x(\vec{r},t)}{\partial t} + \frac{\partial B_z(\vec{r},t)}{\partial y} + \frac{v}{c}\frac{\partial E_x(\vec{r},t)}{\partial y}).
\label{eq:amperex}
\end{eqnarray}
Then, combining Equations (\ref{eq:amperex}) and assuming that $x-$current in $S'$ transforms as $J_x'=J_x$, it is possible to show that  
\begin{equation}
\frac{\partial E_x'(\vec{r}',t')}{\partial t'} = (c\nabla'\times\vec{B}'(\vec{r}',t'))_x -4\pi J_x'.
\label{eq:amperex}
\end{equation} 
The assumption $J_x'=J_x$ will be checked below by considering the way charge density  transforms under Equations \ref{eq:lorentz}.

\subsection{{\bf Ampere's equation: $y$ component}}
Here we proceed similarly as for the $x$ component of the Ampere's equation. From the field transformations defined in Equations (\ref{eq:fields}) we get that
\begin{equation}
\begin{array}{rclcccrclcccrcl}
E'_y(\vec{r}',t') &=& E_y(\vec{r},t),&&&&B'_z(\vec{r}',t') &=& \Gamma (B_z(\vec{r},t) + \frac{v}{c}E_x(\vec{r},t)),&&\textrm{and}&&B'_x(\vec{r}',t') &=& \Gamma (B_x(\vec{r},t) - \frac{v}{c}E_z(\vec{r},t)).
\end{array}
\end{equation}
Thus, from Equations \ref{eq:lorentz} it is possible to show that
\begin{eqnarray}
\frac{\partial E_y'(\vec{r}',t')}{\partial t'} &=& \Gamma\Big(\frac{\partial E_y(\vec{r},t)}{\partial t} + v\frac{\partial E_y(\vec{r},t)}{\partial y}\Big), \nonumber \\
\frac{\partial B_x'(\vec{r}',t')}{\partial z'} &=& \Gamma \Big(\frac{\partial B_x(\vec{r},t)}{\partial z} - \frac{v}{c}\frac{\partial E_z(\vec{r},t)}{\partial z}\Big), \textrm{ and} \nonumber \\ 
\frac{\partial B_z'(\vec{r}',t')}{\partial x'} &=& \Gamma\Big(\frac{\partial B_z(\vec{r},t)}{\partial x} +  \frac{v}{c}\frac{\partial E_x(\vec{r},t)}{\partial x} \nonumber \\&&
+\Big(\frac{\partial \Gamma}{\partial x}(y'+vt') + \Gamma t' \frac{\partial v}{\partial x}\Big)\Big(\frac{\partial B_z}{\partial y} + \frac{v}{c}\frac{\partial E_x}{\partial y}\Big)
+\Big(\frac{\partial \Gamma}{\partial x}(t'+\frac{vy'}{c^2}) + \frac{\Gamma y'}{c^2} \frac{\partial v}{\partial x}\Big)\Big(\frac{\partial B_z}{\partial t} + \frac{v}{c} \frac{\partial E_x}{\partial t}\Big).
\label{eq:amperey}
\end{eqnarray}
Combining Equations \ref{eq:amperey} and assuming $J_y' = \Gamma(J_y -v\rho_c)$ (which we will check below) we obtain
\begin{equation}
\frac{\partial E_y'(\vec{r}',t')}{\partial t'} = (c\nabla'\times\vec{B}'(\vec{r}',t'))_y -4\pi J_y' - sE_x'(\vec{r}',t') - s\Big(ct'\frac{\partial B_z'}{\partial y'} + \frac{y'}{c}\frac{\partial B_z'}{\partial t'}\Big).
\label{eq:amperey}
\end{equation} 

\subsubsection{{\bf Ampere's equation: $z$ component}}
Finally, from the transformations \ref{eq:fields} we get:
\begin{equation}
\begin{array}{rclcccrclcccrcl}
E'_z(\vec{r}',t') &=& \Gamma (E_z(\vec{r},t) - \frac{v}{c}B_x(\vec{r},t)),&&&&B'_x(\vec{r}',t') &=& \Gamma (B_x(\vec{r},t) - \frac{v}{c}E_z(\vec{r},t)),&&\textrm{and}&&B'_y(\vec{r}',t') &=& B_y(\vec{r},t).
\end{array}
\end{equation}
Also, from \ref{eq:lorentz} we obtain
\begin{eqnarray}
\frac{\partial E_z'(\vec{r}',t')}{\partial t'} &=& \Gamma^2 (\frac{\partial E_z(\vec{r},t)}{\partial t} - \frac{v}{c}\frac{\partial B_x(\vec{r},t)}{\partial t} +v\frac{\partial E_z(\vec{r},t)}{\partial y} - \frac{v^2}{c}\frac{\partial B_x(\vec{r},t)}{\partial y}),\nonumber \\
\frac{\partial B_x'(\vec{r}',t')}{\partial y'} &=& \Gamma^2 (\frac{v}{c^2}\frac{\partial B_x(\vec{r},t)}{\partial t} - \frac{v^2}{c^3}\frac{\partial E_z(\vec{r},t)}{\partial t} + \frac{\partial B_x(\vec{r},t)}{\partial y} - \frac{v}{c}\frac{\partial E_z(\vec{r},t)}{\partial y}), \textrm{ and} \nonumber \\ 
\frac{\partial B_y'(\vec{r}',t')}{\partial x'} &=& \frac{\partial B_y(\vec{r},t)}{\partial x} 
+\Big(\frac{\partial \Gamma}{\partial x}(y'+vt') + \Gamma t' \frac{\partial v}{\partial x}\Big)\frac{\partial B_y}{\partial y}
+\Big(\frac{\partial \Gamma}{\partial x}(t'+\frac{vy'}{c^2}) + \Gamma \frac{y'}{c^2} \frac{\partial v}{\partial x}\Big)\frac{\partial B_z}{\partial t}
\label{eq:amperez}
\end{eqnarray}
Combining Equations (\ref{eq:amperez}) and assuming $J_z'=J_z$ (which will be checked below) it is possible to show that  
\begin{equation}
\frac{\partial E_z'(\vec{r}',t')}{\partial t'} = (c\nabla'\times\vec{B}'(\vec{r}',t'))_z -4\pi J_z' +s \Big(ct'\frac{\partial B_y'}{\partial y'} + \frac{y'}{c}\frac{\partial B_y'}{\partial t'}\Big).
\label{eq:amperez}
\end{equation} 
Thus, combining the three components of Ampere's law (Equations \ref{eq:amperex}, \ref{eq:amperey}, and \ref{eq:amperez}) we get
\begin{equation}
\frac{\partial \vec{E}'(\vec{r}',t')}{\partial t'} = c\nabla'\times\vec{B}'(\vec{r}',t') -4\pi\vec{J}'-sE'_x(\vec{r}',t')\hat{y}  - s\Big(ct'\frac{\partial \vec{B}'(\vec{r}',t')}{\partial y'} + \frac{y'}{c}\frac{\partial \vec{B}'(\vec{r}',t')}{\partial t'}\Big)\times \hat{x}.
\label{eq:ampere}
\end{equation}
As with Faraday's equation, Ampere's equation also gets a time dependent term that disappears under the 2D approximation. Similarly, the 2D limit requires the $y'$ term to be negligible so that the equations governing the evolution of $\vec{E}'$ are independent of $y'$. As in the case of Faraday's equation, this can be done by noticing that, since $\partial \vec{B'}/\partial t' \approx \nabla' \times \vec{E'}$, then $\nabla' \times \vec{B}' \gg sy'\partial \vec{B}'/\partial t'$, provided that $sy' \ll c$ and $|\vec{E}'| \ll |\vec{B}'|$. This means that the $y'-$dependent term $sy'\partial \vec{B}'/\partial t'$ can also be neglected in the case of Ampere's equation. It is important to notice, however, that $sy'\partial \vec{B}'/\partial t'$ is not necessarily much smaller than $\partial \vec{E}'/\partial t'$. Thus, although neglecting the $y'-$dependent term does not modify the MHD-scale dynamics (e.g., that of the MRI), it does not fully capture the evolution of the curl-less electric field component. Thus, our approach neglects the appearance of electric charges that in principle could influence the microphysics of the plasma. These charges, however, appear to be smaller than the ones due to frame rotation (see discussion in \S \ref{sec:baseq}), which we are already neglecting in the context of the kinetic MRI dynamics. \newline

Thus, assuming a 2D geometry, Ampere's equation becomes:
\begin{equation}
\frac{\partial \vec{E}'(\vec{r}',t')}{\partial t'} = c\nabla'\times\vec{B}'(\vec{r}',t') -4\pi \vec{J}' - sE_x'(\vec{r}',t')\hat{y}.
\label{eq:amperetotal}
\end{equation}

\subsection{{\bf Particle Momentum Evolution In The $S'$ Frame.}}
The modified version of Ampere's equation assumes that currents transform as
\begin{equation}
\begin{array}{rclcccrclcccrcl}
J_x'&=&J_x,&\textrm{  }&\textrm{  }&\textrm{  }&J_y'&=&\Gamma(J_y-v\rho_c), &\textrm{  }&\textrm{ and }&\textrm{  }&J_z'&=&J_z.
\end{array}
\label{eq:current}
\end{equation}
Based on the coordinate transformation \ref{eq:lorentz}, it is possible to show that the charge density in $S'$ ($\rho'_{c}$) is related to the one in $S$ by
\begin{equation}
\rho'_c=\Gamma(\rho_c -vJ_y/c^2+sy'J_x/(c^2\Gamma)),
\label{eq:denstrans}
\end{equation}
where $\vec{u}$ is the fluid velocity in $S$.
Equation \ref{eq:denstrans} can be simplified under the assumptions $v, sy' \ll c$, implying that, in the non-relativistic limit, charge densities do not transform. This means that simple Galilean velocity transformations are enough to reproduce the non-relativistic version of current transformation equations \ref{eq:current}. However, since individual particles in our simulations are allowed to become relativistic, we will use the relativistic transformation of their momenta, $\vec{p}$. This way, we ensure that no particles in $S'$ exceed the speed of light. Thus,
\begin{equation}
\begin{array}{rclcccrcl}
p_x'&=&p_x,&\textrm{  }&\textrm{  }&\textrm{  }&p_y'&=&\Gamma(p_y-v\gamma),\\
p_z'&=&p_z,&\textrm{  }&\textrm{ and }&\textrm{  }&\gamma'&=&\Gamma(\gamma -vp_y/(mc^2)),
\end{array}
\label{eq:mom}
\end{equation}
where $m$ is the mass of the particle. These relations (along with the time transformation shown in Equation \ref{eq:lorentz}) imply that
\begin{eqnarray}
\frac{dp_x'}{dt'}&=&\frac{dp_x/dt}{\Gamma(1-vu_y/c^2)+su_xy'/c^2}, \nonumber \\
\frac{dp_y'}{dt'}&=& \frac{dp_y/dt - mvd\gamma/dt +sp_x(1-vu_y/c^2)}{1-vu_y/c^2+su_xy'/(c^2\Gamma)}, \textrm{     and} \nonumber \\
\frac{dp_z'}{dt'}&=&\frac{dp_z/dt}{\Gamma(1-vu_y/c^2)+su_xy'/c^2},
\label{eq:translorentz}
\end{eqnarray} 
where $u_x$ and $u_y$ are the $x$ and $y$ velocities of the particle in the $S$ frame. It is straightforward to show that, in the non-relativistic limit ($v, sy' \ll 1$), Equations \ref{eq:translorentz} correspond to the standard force transformation between two inertial frames, plus an extra force $sp'_x\hat{y}$. Thus, given that the electric and magnetic fields transform analogous to the usual relativistic field transformation, one can easily show that the time evolution of the particles momenta in $S'$ will be given by the Lorentz forces, plus the extra term $sp'_x\hat{y}$.\newline
In addition to these forces, we also need to find out what is the transformation of the coriolis and tidal forces forces on the particles. We know that in the $S$ frame\footnote{This expression is valid in the cold limit, i.e., $|\vec{u}| \ll |\vec{v_0}|$. Since in our simulations individual particles are allowed to get accelerated to velocities $|\vec{u}| \gg |\vec{v_0}|$, the validity of the cold limit will be in a fluid sense. Thus, as long as the mean velocities of the particles is non-relativistic, the contribution of gravity to particles dynamics will be well described by Equation \ref{eq:cortidal}.}
\begin{equation}
\frac{d\vec{u}}{dt} = 3\omega_0^2x \hat{x} - 2\vec{\omega}_0 \times \vec{u}.
\label{eq:cortidal}
\end{equation}
It is straightforward to show that, in the cold limit, the $S'$ version of Equation \ref{eq:cortidal} becomes
\begin{equation}
\frac{d\vec{u}'}{dt} = -2\omega_0 \hat{z}\times \vec{u}'. 
\label{eq:cortidalsh}
\end{equation}
Since this force does not do work on a particle, it can be equivalently expressed in terms of $\vec{p}'$ instead of $\vec{u}'$. Thus, combining Equation \ref{eq:cortidalsh} with the transformation to Lorentz forces found above, we get 
\begin{equation}
\frac{d\vec{p}'}{dt} = 2\omega_0 p_y'\hat{x} - \frac{1}{2}\omega_0 p_x'\hat{y} +q(\vec{E}'+\frac{\vec{u}'}{c}\times\vec{B}'),
\label{eq:finalforce}
\end{equation} 
where $q$ is the charge of the particle.\newline
Finally, it is important to point out that the momentum transformation shown in Equations \ref{eq:mom} is not entirely consistent with the evolution of particles positions. Indeed, by directly differentiating Equation \ref{eq:lorentz} one gets
\begin{eqnarray}
u_x'&=
&\frac{u_x}{\Gamma(1-vu_y/c^2)+su_xy'/c^2}, \nonumber
\\
u_y'&=&\frac{u_y - v + su_xt'/\Gamma}{(1-vu_y/c^2)+su_xy'/(c^2\Gamma)}, \textrm{   }\textrm{ and} \nonumber
\\
u_z'&=&\frac{u_z}{\Gamma(1-vu_y/c^2)+su_xy'/c^2}.
\label{eq:velos}
\end{eqnarray}
In the non-relativistic regime, the discrepancy appears only in $u'_y$. In the 2D case, this does not imply an inconsistency between the values of $u'_y$ and the evolution of the $y'$ position of particles (where by definition $y'=0$). In 3D, on the other hand, this implies a violation of charge conservation, which will need to be taken into account in future 3D studies of this problem.

\label{sec:appendix}


\begin{thebibliography}{99}
\addcontentsline{toc}{section}{Bibliography}

\bibitem[Balbus \& Hawley(1991)]{BalbusEtAl91} Balbus, S. A., \& Hawley, J. F. 1991, \apj, 376, 214
\bibitem[Balbus (2004)]{BalbusEtAl04} Balbus, S. A. 2004, \apj, 616, 857
\bibitem[Balbus \& Hawley(1998)]{BalbusEtAl98} Balbus, S. A., \& Hawley, J. F. 1998, Rev. Mod. Phys., 70, 1
\bibitem[Bale et al.(2009)]{BaleEtAl09} Bale, S. D., Kasper, J. C., Howes, G. G., Quataert, E., Salem, C., \& Sundkvist, D. 2009, Phys. Rev. Lett. 103, 211101
\bibitem[Buneman(1993)]{Buneman93} Buneman, O. 1993, ``Computer Space Plasma Physics'', Terra Scientific, Tokyo, 67
\bibitem[Dexter et al.(2010)]{DexterEtAl10} Dexter, J., Agol, E, Fragile, P. C., \& McKinney, J. C. 2010, \apj, 717, 1092
\bibitem[Drake et al.(2006)]{DrakeEtAl06} Drake, J. F., Swisdak, M., Che, H., \& Shay, M. A. 2006, Nature, 443, 553
\bibitem[Esin et al.(1997)]{EsinEtAl97} Esin, A. A., McClintock, J. E., \& Narayan, R. 1997, \apj, 489, 865
\bibitem[Ferraro(2007)]{Ferraro07} Ferraro, N. M. 2007, \apj, 662, 512
\bibitem[Gary et al.(1997)]{GaryEtAl97} Gary, S. P., Wang, J., Winske, D., \& Fuselier, S. A. 1997, J. Geophys. Res., 102, 27159
\bibitem[Hasegawa(1969)]{Hasegawa69} Hasegawa, A. 1969, Phys. Fluids, 12, 2642
\bibitem[Hellinger et al.(2006)]{HellingerEtAl06} Hellinger, P., Travnciek, P. M., Kasper, J. C., \& Lazarus, A. 2006, Geophys Res Lett, 33, L09101
\bibitem[Islam \& Balbus(2005)]{IslamEtAl91} Islam, T., \& Balbus, S. A. 2005, \apj, 633, 328
\bibitem[Johnson et al.(2008)]{JohnsonEtAl08} Johnson, B. M., Guan, X., \& Gammie, C. F. 2008, \apjs, 177, 373
\bibitem[Krolik \& Zweibel(2006)]{KrolikEtAl06} Krolik, J. H., \& Zweibel, E. 2006, \apj, 644, 651
\bibitem[Narayan et al.(1998)]{Narayan1998} Narayan, R., Mahadevan, R., \& Quataert, E.\ 1998, Theory of Black Hole Accretion Disks, 148 
\bibitem[Quataert et al.(2002)]{QuataertEtAl02} Quataert, E., Dorland, W., \& Hammett, G. W. 2002, \apj, 577, 524
\bibitem[Quataert 
\& Gruzinov(1999)]{QG1999} Quataert, E., \& Gruzinov, A.\ 1999, \apj, 520, 248 
\bibitem[Schiff(1939)]{Schiff39} Schiff, L. I. 1939, Proc. Nat. Acad. Sci, 25, 391
\bibitem[Schlickeiser \& Skoda(2010)]{SchlickeiserEtAl10} Schlickeiser, R. \& Skoda, T. 2010, \apj, 716, 1596
\bibitem[Shakura \& Sunyaev(1973)]{ShakuraEtAl73} Shakura, N. I., Sunyaev, R. A. 1973, \aa, 24, 337
\bibitem[Sharma et al.(2003)]{SharmaEtAl03} Sharma, P., Hammett, G. W., \& Quataert, E. 2003, \apj,
596, 1121
\bibitem[Sharma et al.(2006)]{SharmaEtAl06} Sharma, P., Hammett, G. W., Quataert, E., \& Stone J. M. 2006, \apj, 637, 952
\bibitem[Sharma et al.(2007)]{SharmaEtAl07} Sharma, P., Quataert, E., Hammett, G. W., \& Stone J. M. 2007, \apj, 667, 714
\bibitem[Sironi \& Spitkovsky(2011)]{SironiEtAl11} Sironi, L., \& Spitkovsky, A. 2011, \apj, 741, 39
\bibitem[Spitkovsky(2005)]{Spitkovsky05} Spitkovsky, A. 2005, AIP Conf. Proc, 801, 345, astro-ph/0603211
\end{thebibliography}
\end{document}